\documentclass[aps,pra,superscriptaddress,twocolumn,10pt]{revtex4-1}
\usepackage[colorlinks,allcolors=blue]{hyperref}
\usepackage{amsmath}
\usepackage{amsfonts}
\usepackage{wasysym}
\usepackage{newtxtext,newtxmath}
\usepackage{color}
\usepackage{pict2e}
\usepackage{graphicx}
\usepackage{mathrsfs}
\usepackage{enumitem}
\newcommand{\La}{\line(1,0){12}}
\newcommand{\Lb}{\line(3,5){6}}
\newcommand{\Lc}{\line(-3,5){6}}
\newcommand{\Ld}{\line(-1,0){12}}
\newcommand{\Le}{\line(-3,-5){6}}
\newcommand{\Lf}{\line(3,-5){6}}
\newcommand{\CC}{\circle*{4}}

\newcommand{\pA}{\put(-6,-10)}
\newcommand{\pB}{\put(6,-10)}
\newcommand{\pC}{\put(12,0)}
\newcommand{\pD}{\put(6,10)}
\newcommand{\pE}{\put(-6,10)}
\newcommand{\pF}{\put(-12,0)}

\newcommand{\Hex}{\pA{\CC}\pB{\CC}\pC{\CC}\pD{\CC}\pE{\CC}\pF{\CC}}
\newcommand{\ha}{
    \begin{picture}(32,18)(-16,-5)
        \Hex\pA{\La}\pC{\Lc}\pE{\Le}
    \end{picture}}
\newcommand{\hb}{
    \begin{picture}(32,18)(-16,-5)
        \Hex
        \pB{\Lb}\pD{\Ld}\pF{\Lf}
    \end{picture}}

\begin{document}

\title{Emergent Rokhsar-Kivelson point in realistic quantum Ising models}
\author{Zheng Zhou}
\affiliation{Department of Physics and State Key Laboratory of Surface Physics, Fudan University, Shanghai 200438, China}
\author{Zheng Yan}
\affiliation{Department of Physics and HKU-UCAS Joint Institute of Theoretical and Computational Physics, The University of Hong Kong, Pokfulam Road, Hong Kong}
\affiliation{Department of Physics and State Key Laboratory of Surface Physics, Fudan University, Shanghai 200438, China}
\author{Changle Liu}
\email{liuchangle89@gmail.com}
\affiliation{Shenzhen Institute for Quantum Science and Technology and Department of Physics, Southern University of Science and Technology, Shenzhen 518055, China}
\affiliation{Department of Physics and Center of Quantum Materials and Devices, Chongqing University, Chongqing, 401331, China}
\author{Yan Chen}
\affiliation{Department of Physics and State Key Laboratory of Surface Physics, Fudan University, Shanghai 200438, China}
\affiliation{Collaborative Innovation Center of Advanced Microstructures, Nanjing 210093, China}
\author{Xue-Feng Zhang}
\affiliation{Department of Physics and Center of Quantum Materials and Devices, Chongqing University, Chongqing, 401331, China}
\date{\today}

\begin{abstract}
    We show that the Rokhsar-Kivelson~(RK) point in quantum dimer models~(QDM) can emerge in realistic quantum Ising spin systems. Specifically, we investigate the $J_1$-$J_2$-$J_3$ transverse field Ising model on the triangular lattice with large-scale quantum Monte Carlo simulations. We find that the multicritical point in the phase diagram corresponds to the RK point of the QDM on the honeycomb lattice. We further measure the spectral functions and identify three branches of quadratic dispersions. In the phase diagram, we also find a sequence of incommensurate states, which meet the microscopic ingredients for the `Cantor deconfinement' scenario. Our study provides a promising direction to realise the RK deconfinement in experimental platforms such as magnetic materials or programmable Rydberg arrays.
\end{abstract}

\maketitle

\emph{Introduction.}---Searching for exotic states of matter has been a central task in the physics community. In the past decades, a series of novel quantum states have been discovered in which correlations among elementary degrees of freedom are not manifested by means of broken symmetries, but more `hidden' features such as fractionalised excitations, emergent gauge structures, and long-range quantum entanglement~\cite{wen2017zoo,zhou2017quantum,castelnovo2008magnetic,Balents,BalentsSavary,horodecki2009quantum}. These exotic features can be clearly understood in systems with locally constrained Hilbert space~\cite{Balents,Hermele,BFG,lgt_review}. A paradigmatic example is the quantum dimer model~(QDM)~\cite{rk_debut,qdm_review}, in which the Hilbert space is restricted to fully-packed dimer coverings such that each site is joint with one and only one dimer. The original QDM was proposed by Rokhsar and Kivelson~(RK) on the square lattice~\cite{rk_debut}, and it takes the form on the honeycomb lattice as
\begin{equation}
    H_\mathrm{RK}=\sum_{\hexagon} v\left(|\ha\rangle\langle\ha|+|\hb\rangle\langle\hb|\right)-t\left(|\ha\rangle\langle\hb|+\mathrm{H.c.}\right).
    \label{eq:RKQDM}
\end{equation}
The QDM is proposed to capture the low energy fluctuations of valence bond systems~\cite{rk_debut,qdm_vbs}. Meanwhile, it also arises in certain limits of some frustrated Ising models~\cite{moessner01,frus_ising_1,frus_ising_2,zhou_string,fradkin_cmft,moessner_2000,yan2021sweeping}. QDMs not only provide a natural implementation of lattice gauge theories with various gauge structures~\cite{LGT_QDM,lgt_review}, but also offer valuable understandings for various exotic behaviours in frustrated magnets.

\begin{figure}[ht!]
    \centering
    \includegraphics[width=1\linewidth]{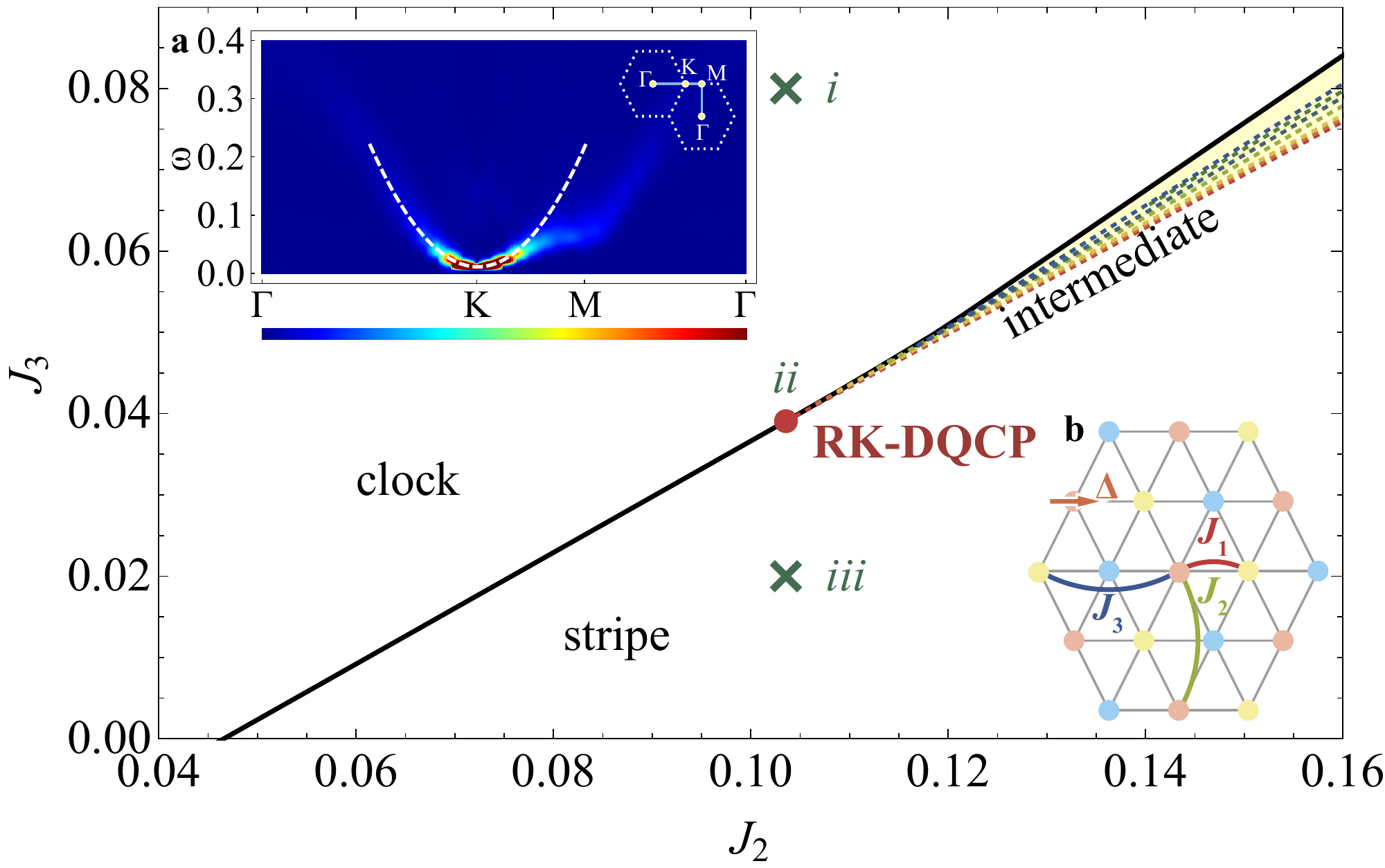}
    \caption{Ground state phase diagram of the $J_1$-$J_2$-$J_3$ TFIM. Bold black lines mark first-order transitions. A fan of intermediate phases with `tilt' $0<f<2$ are separated by coloured lines. The green crosses labeled \textit{i}--\textit{iii} mark the points at which the histogram of plaquette VBC order parameter $\phi$ is measured~(Fig.~\ref{fig:3}). Insets: (a) The quadratic dispersion in the spin-spin correlation spectrum at the multicritical point of the $J_1$-$J_2$-$J_3$ TFIM. The numerical simulations are performed with system size $L=24$. (b) Definition of the triangular lattice and the interactions of the model Eq.~(\ref{eq:j123}). Three sublattices are marked yellow, blue and pink, respectively.}
    \label{fig:1}
\end{figure}

The QDMs exhibit interesting physics at an exactly soluble point $v=t$, dubbed as the `RK point'~\cite{rk_debut,qdm_vbs}. Remarkably, this point corresponds to deconfined quantum liquid states, providing the first microscopic realisation of Anderson's resonating valence bond~(RVB) proposal~\cite{ANDERSON_RVB,AndersonFazekas}.
On 2$d$ bipartite lattices, the RK point turns out to be a deconfined quantum critical point~(DQCP) ~\cite{senthil2004deconfined} separating different symmetry-breaking crystalline phases~\cite{rk_2dbp}.
More precisely, the continuous transition occurs at only one side of the critical point, and across which the system immediately exhibit sharp discontinuity~\cite{rk_2dbp,qdm_review}. This behaviour is somewhat different from the N\'eel-to-VBC DQCP~\cite{senthil2004deconfined} that people are more familiar with.
Recently, RK-DQCP has been unveiled with non-trivial correlation and entanglement properties~\cite{0812_0203,0906_1569}. Moreover, perturbations around RK-DQCP can stabilise a sequence of commensurate and incommensurate phases in the phase diagram with fractal `devil's staircase' structure~\cite{rk_2dbp,qdm_honey}. These incommensurate states exhibit gapless phason excitations~\cite{phason} which turn out to deconfine monomers, providing a new route to deconfinement~\cite{rk_2dbp,ashvin1}. However, despite the remarkable properties of RK-DQCP, its experimental realisation remains a challenging issue until recently~\cite{zoller2014twodimensional,qdm_real,zoller}.

In this manuscript, we show that the RK-DQCP in honeycomb QDMs can be realised in triangular lattice Ising spin systems that are relevant to experiments. We first briefly review the field theory around RK-DQCP and show that the RK-DQCP on generic honeycomb QDMs has two relevant perturbations. After establishing connections between the honeycomb QDM and the triangular TFIM, we show that these two relevant perturbations can be manipulated by two additional neighboured spin interactions. To be concrete, we numerically investigate the antiferromagnetic $J_1$-$J_2$-$J_3$ transverse field Ising model~(TFIM) with large-scale quantum Monte Carlo~(QMC) simulations. We find that the multicritical point in the phase diagram corresponds to the RK-DQCP in honeycomb QDMs. In particular, we have observed quadratic dispersions in the spectra which signifies the dynamical exponent $z=2$~(Fig.~\ref{fig:1}a) and an intermediate regime associated with Cantor's deconfinement in the phase diagram~(Fig.~\ref{fig:1}). We further discuss the experimental relevance of our work with condensed matter and cold atom systems.

\emph{Field theory around RK-DQCP.}---On 2$d$ bipartite lattices, the local constraints of the dimer Hilbert space yields an emergent $U$(1) gauge structure~\cite{LGT_QDM,yan_mixed}: each dimer configurations can be mapped onto the compact height scalar field $\tilde h$ on the dual lattice as the $U$(1) gauge potential~\cite{note_supp}. A natural consequence is that dimer coverings can be classified into topological sectors labeled by two integers ($F_1,F_2$) as the winding of $\tilde h$ on two non-contractable loops of the $L_1\times L_2$ periodic lattice~\cite{note_supp}. For the ground states one of the two integers is zero, thus we can set $F_2=0$ and work with $f\equiv F_1/L_1$. While the value of $f$ characterises different topological sectors, it is also proportional to the gradient of the coarse-grained height field $h$, therefore is also referred to as `tilt' in literatures.

The effective field theory in the vicinity of RK-DQCP takes the form~\cite{Henley,qdm_vbs}
\begin{equation}
    \mathscr{L}_0=\frac{1}{2}\left[(\partial_\tau h)^2+\kappa^2(\nabla^2h)^2\right]+\frac{\rho_2}{2}(\nabla h)^2+\lambda\cos2\pi h,
    \label{eq:L0}
\end{equation}
where $\rho_2\propto 1-v/t$ controls the phase transition, and $\lambda$ is the instanton term dictating the discrete nature of the height variable. When $\rho_2>0$, the instanton term $\lambda$ is relevant so that the system is pinned to the three-fold plaquette state with $f=0$; by contrast, the fluctuations of $\nabla h$ immediately become unbounded when $\rho_2<0$, driving the system to the staggered state with saturated `tilt' $f=2$. At the critical point $\rho_2=0$, the instanton term $\lambda$ becomes dangerously irrelevant~\cite{shao2020dangerous,oshikawa2000ordered}, which signals deconfinement. This very point corresponds to the RK-DQCP which is described by the quantum Lifshitz model~\cite{qdm_lifshitz}
\begin{equation}
\mathscr{L}_\mathrm{QLM}=\frac{1}{2}\left[(\partial_\tau h)^2+\kappa^2(\nabla^2h)^2\right].
\label{eq:LQLM}
\end{equation}
As a result, RK-DQCP is a deconfined $U(1)$ liquid state with dynamical exponent $z=2$, free from Polyakov's argument that pure compact quantum electrodynamics in $(2+1)\mathrm{d}$ is always confining~\cite{Polyakov}.

More recently, the fate of RK-DQCP in the presence of generic perturbations has been discussed in Refs.~\cite{rk_2dbp,ashvin1}. On the honeycomb lattice, it was argued that RK-DQCP also admit a relevant trigonal anisotropy $\mathscr{L}_1=g_3\prod_{\alpha=1}^{3}\left(\nabla h\cdot\hat{\mathbf{e}}_\alpha\right)$ in addition to a marginally irrelevant quartic coupling $\mathscr{L}_2=g_4\left(\nabla h\cdot\nabla h\right)^2$. Here $\hat{\mathbf{e}}_\alpha$ ($\alpha=1,2,3$) are the unit vectors aligned perpendicular to the three dimer directions. Considering the whole Lagrangian
\begin{equation} 
\mathscr L=\mathscr L_0 +\mathscr L_1+\mathscr L_2,
\label{eq:L012}
\end{equation}
 it was shown that a sequence of commensurate and incommensurate regime with finite but non-saturated $f$ is stabilised in the vicinity of the RK-DQCP. This intermediate regime is argued to form an incomplete `devil's staircase' structure, with the analog of the fractal `Cantor set' in mathematics. The presence of gapless phason mode in the incommensurate phase corresponds to gapless photons that prevent monomers from deconfining in the $U(1)$ gauge theory. This scenario is dubbed as `Cantor deconfinement' as proposed in the seminal work~\cite{rk_2dbp}.

\begin{figure}[t]
    \centering
    \includegraphics[width=1\linewidth]{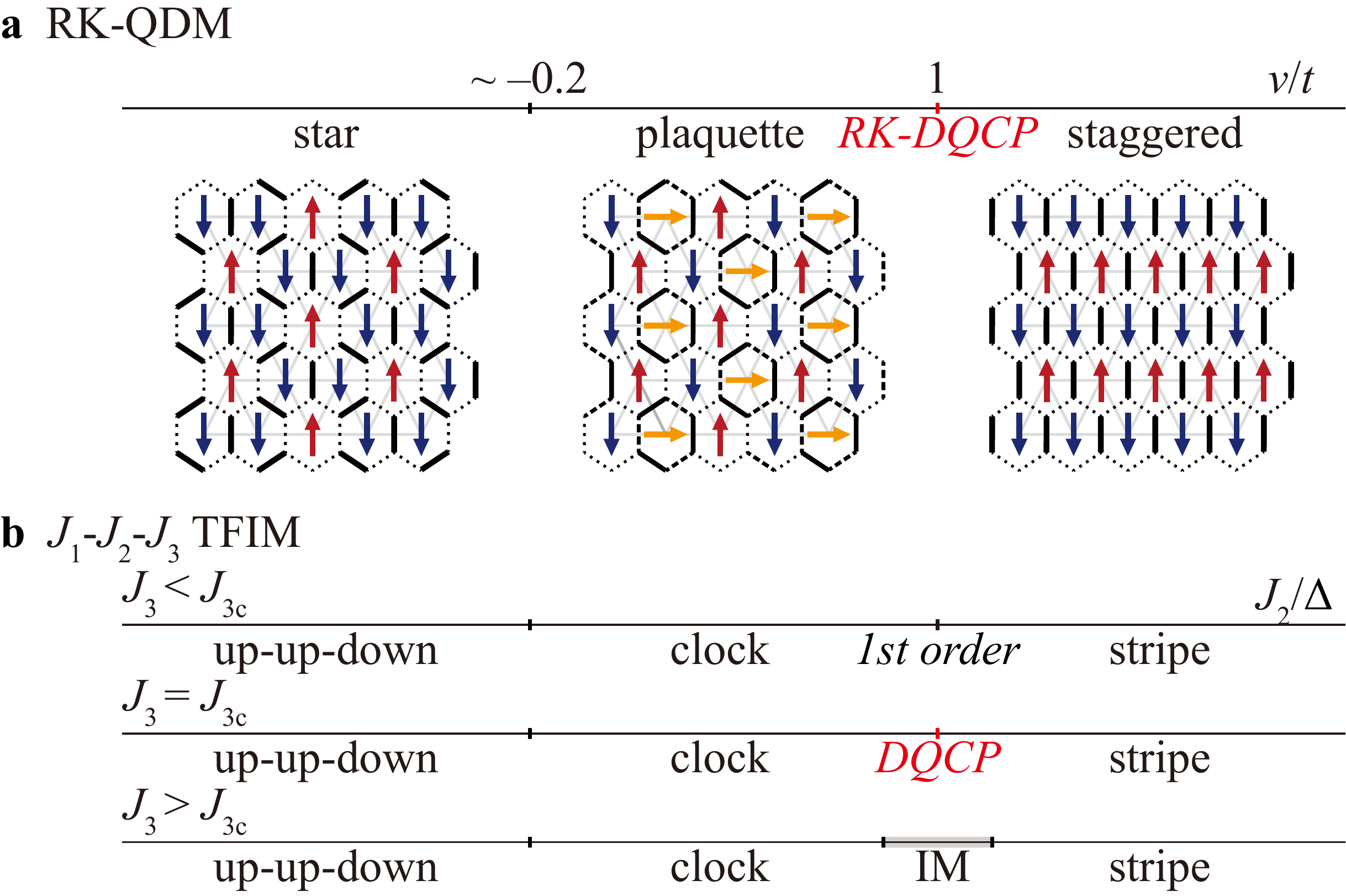}
    \caption{(a) The phase diagram of honeycomb RK-QDM, with each phase and their mappings to spin configurations illustrated. (b) The phase diagram of triangular $J_1$-$J_2$-$J_3$ TFIM with $J_2$ as the tuning parameter at different $J_3$ (The $J_1$-$J_2$ model also falls into the $J_3<J_{3c}$ catagory). The `IM' stands for the intermediate phase (the spin configurations are not drawn here). }
    \label{fig:2}
\end{figure}

\emph{Dimer-to-spin mapping.}---The connection between honeycomb QDM and triangular TFIM was first established in Refs.~\cite{moessner01,frus_ising_2}. Here we briefly describe the mapping process: we define Ising spin-$1/2$ objects at the center of each honeycomb hexagon~(Fig.~\ref{fig:2}a); the spins then form a triangular lattice; the Ising spins are aligned such that each triangle-lattice bond that crosses a dimer connects parallel spins, and otherwise connects anti-parallel spins. The one-dimer-per-site constrait is recovered by introducing a large nearest-neighbour (NN) antiferromagnetic interaction $J_1\sum_{\langle ij\rangle_1} S_i^z S_j^z$, which guarantees that exactly one frustrated bond appears within each triangle unit. Treating the transverse field term $\left(-\Delta\sum_{i}S_{i}^{x}\right)$ as a  perturbation, we find that in the low energy dimer manifold it translate into the dimer flipping term $t$ in Eq.~(\ref{eq:RKQDM}) with $\Delta=t/2$. The above mapping establishes the connection between the $J_1$ TFIM and the RK-QDM with $v=0$, $t=2\Delta$. Note that the quantum fluctuation $\Delta$ takes effect in the low energy dimer manifold at the first order of the perturbation theory. It implies significant gauge fluctuations compared with quantum spin ice systems, where one needs to go to at least third-order to obtain a dimer flipping term~\cite{Hermele,Savary}. It should also be noticed that the mapping from dimer to spin manifold is redundant~\cite{moessner01,qdm_honey}: each dimer configuration $c$ corresponds to two spin configurations $|\phi_{c\pm}\rangle$ related by a global $\mathbb{Z}_2$ flipping $\mathscr G=\prod_i2S_i^x$. The dimer-to-spin redundancy brings non-trivial consequences, one of which is that any physical operator defined within the dimer Hilbert space must not be affected by such redundancy, therefore must stay invariant under $\mathscr G$.

One major obstacle against the experimental realisation of RK-QDM is the artificiality of the diagonal $v$ term in Eq.~(\ref{eq:RKQDM}). This term counts the overall number of flippable plaquettes, which involves unrealistic multiple spin interactions~\cite{qdm_honey,yan_sweeping,yan_sweeping_2} in the spin representation. Therefore it is greatly appreciated to substitute the $v$ term with pairwise spin interactions realistic in nature. Interestingly, we observe that the next-NN interaction $J_2\sum_{\langle ij\rangle_2}S_i^z S_j^z$ plays a similar role to the $v$ term in Eq.~(\ref{eq:RKQDM}). This similarity is illustrated by comparing the phase diagram of RK-QDM and the $J_1$-$J_2$ TFIM in the limit $J_1\gg J_2, \Delta$~(Fig.~\ref{fig:2}a)~\cite{Damle16,zhou2020quantum}. Strikingly, we observe an almost exact correspondence between the phase diagram of RK-QDM and that of the $J_1$-$J_2$ TFIM. The only exception is that the clock-to-stripe transition in the TFIM is trivially first-order~\cite{note_1}, in contrast to a non-trivial DQCP that appears in the RK-QDM counterpart.

The distinct nature of phase transition indicates that the RK-QDM and $J_1$-$J_2$ TFIM are not completely equivalent. The underlying reason is that the $J_2$ term not only plays the role of the $v$ term that tunes $\rho_2$, but also inevitably induces some trigonal anisotropy $g_3$. We notice that, however, the relevant perturbation $g_3$ can be eliminated by further introducing a third-NN interaction $J_3\sum_{\langle ij\rangle_3}S_i^zS_j^z$. A more detailed discussion is given in Supplemental Material~\cite{note_supp}. As RK-DQCP has only two relevant perturbations, we expect RK-DQCP to appear in the phase diagram once we consider both $J_2$ and $J_3$.

\emph{Model.}---To validate our expectations, we numerically investigate the $J_1$-$J_2$-$J_3$ TFIM
\begin{equation}
H=\sum_{n=1}^{3}J_{n}\sum_{\langle ij\rangle_n}S_i^zS_j^z-\Delta\sum_iS_i^x
\label{eq:j123}
\end{equation}
where $\langle ij\rangle_n$ denotes the $n$-th-NN bonds with large scale quantum Monte-Carlo simulations~\cite{qmc_1,qmc_2,qmc_3,note_supp}. The simulations are performed on $L\times L$ lattices ($L=24,36$) with periodic boundary condition, and the temperature is set to $T=L^{-2}$. For convenience and without loss of generality, we set $J_1=1$ as the unit and fix $\Delta =0.2$. The numerical phase diagram is shown in Fig.~\ref{fig:1}. With small $J_2$ and $J_3$, the transition between the clock and the stripe phase is trivially first-order~\cite{note_1}. When $J_2$ and $J_3$ exceeds some critical value, a fan of states with intermediate $f$ emerges at the clock-stripe boundary~(Fig.~\ref{fig:2}b). The intermediate regime is separated with the clock phase through a first-order phase transition and with the stripe phase through a continuous one. The clock-stripe transition line and the fan-shaped region terminate at a single multicritical point. The position of the multicritical point is determined by the intersection between the clock-to-stripe and the stripe-to-intermediate transition lines. Through a finite size scaling analysis, we extrapolate its position in thermodynamic limit to be $J_{2c}=0.1093(11)$ and $J_{3c}=0.0429(7)$~\cite{note_supp}. The structure of the phase diagram resembles that of Eq.~(\ref{eq:L012})~\cite{qdm_honey}, which suggests that the multicritical point in the phase diagram corresponds to the RK-DQCP described by Eq.~(\ref{eq:LQLM}). In the following, we present numerical evidences on the RK-DQCP nature of this multicritical point.

\emph{Degeneracy of topological sectors}.---The effective action of the RK-DQCP Eq.~(\ref{eq:LQLM}) exhibits no dependence on $\nabla h$, therefore the ground states within different topological sectors are exactly degenerate. This degeneracy is a characteristic feature of RK points that can be checked numerically. We measure the ground state energies within different topological sectors at the multicritical point, and find vanishing energy differences between different topological sectors $\sim6\times10^{-4} L^{2}\Delta$~\cite{note_3}. The tiny energy difference may be accounted by the presence of marginal or dangerously irrelevant terms, which remains with finite system size. The proximate degeneracy provide us with a strong hint that the multicritical point corresponds to the RK-DQCP. 

\begin{figure}[t]
    \centering
    \includegraphics[width=1\linewidth]{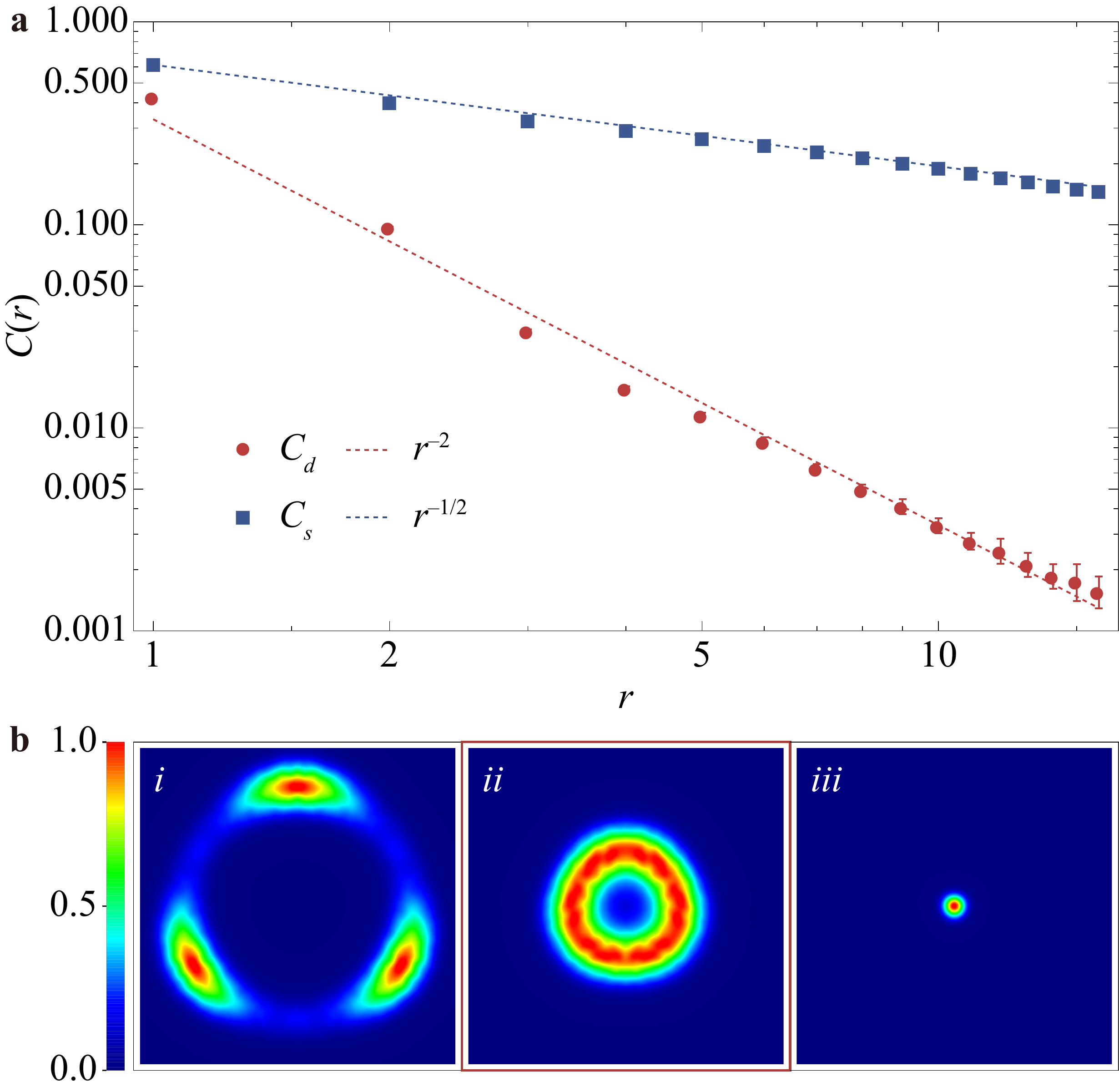}
    \caption{(a)~Dimer-dimer and spin-spin correlators $C_d(r)$ and $C_s(r)$ at the multicritical point. Our numerical results well agree with the theoretical predictions $C_d\sim r^{-2}$ and $C_s\sim r^{-1/2}$ indicated by dashed lines, respectively. The system size in the measurement is $L=36$. (b)~Histograms of the VBC order parameter $\phi$ measured at representative points in Fig.~\ref{fig:1}. The system size of measurement is $L=24$. }
    \label{fig:3}
\end{figure}

\emph{Correlations.}---To search for more evidence of RK-DQCP, we examine the correlation behaviours at the multicritical point. We first measure the dimer-dimer correlator $C_d(\vec{\mathsf{R}}-\vec{\mathsf{R}}')=\langle\phi^*(\vec{\mathsf{R}})\phi(\vec{\mathsf{R}}')\rangle$, where
\begin{equation}
    \phi=S_B^zS_C^z+S_C^zS_A^z\mathrm e^{-\mathrm i2\pi/3}+S_A^zS_B^z\mathrm e^{\mathrm i2\pi/3}
    \label{VBC_OP}
\end{equation}
is the plaquette valence bond crystal~(VBC) order parameter of dimers. Here $A$, $B$, $C$ correspond to three sublattices of the triangular lattice shown in Fig.~\ref{fig:1}b. At RK-DQCP, the long distance behaviour of the dimer-dimer correlator is predicted to be $C_d(\vec{\mathsf{R}})\sim|\vec{\mathsf{R}}|^{-2}$~\cite{note_supp,dimer_cor}. Such behaviour is clearly observed in our simulations~(Fig.~\ref{fig:3}a, red line). With regard to the spin degree of freedom, we consider the spin-spin correlator $C_s(\vec{\mathsf{R}}-\vec{\mathsf{R}}')=\langle \psi^*(\vec{\mathsf{R}}) \psi(\vec{\mathsf{R}}') \rangle$, where
\begin{equation}
    \psi=S_A^z+S_B^z\mathrm e^{\mathrm i2\pi/3}+S_C^z\mathrm e^{-\mathrm i2\pi/3}
\end{equation}
is the clock order parameter of spins. Notice that the spin-spin correlator is equivalent to the non-local dimer-string correlation in the dimer representation~\cite{note_supp}. The spin-spin correlator at RK-DQCP is predicted to satisfy the asymptotic form $C_s(\vec{\mathsf{R}})\sim|\vec{\mathsf{R}}|^{-1/2}$~\cite{note_supp,tri_cor}, also consistent with our numerical results~(Fig.~\ref{fig:3}a, blue line).

\emph{Emergent $U(1)$ symmetry.}---The deconfined nature of this multicritical point is a result of the irrelevant instanton term $\lambda$ in Eq.~(\ref{eq:L0}), which can be verified in the histogram of the plaquette VBC order parameter $\phi$. In the plaquette phase, $\phi$ is related to the height field by the relation $\phi\sim \exp \left( 2\pi ih/3 \right)$. The irrelevance of $\lambda$ leads to an emergent $U(1)$ symmetry generated by $\phi\rightarrow\phi e^{i\theta}$ where $\theta$ is an arbitrary angle. To identify the emergent $U(1)$ symmetry, we examine the histogram of $\phi$ in vicinity to the multicritical point. The result is shown in Fig.~\ref{fig:3}b. In the plaquette ordered state, $\phi$ is pinned to three distinct values in the histogram, indicating the relevance of the $\lambda$ cosine term~(Fig.~\ref{fig:3}b,i). By contrast, the staggered state does not acquire any three-sublattice ordering, therefore the corresponding histogram shows a central peak~(Fig.~\ref{fig:3}b,iii). On approaching the multicritical point, the $U(1)$ symmetry emerges as a symmetric ring in the histogram~(Fig.~\ref{fig:3}b,ii). The right presence of emergent $U(1)$ symmetry at the multicritical point is in full consistency with the deconfined nature associated with the irrelevance of the instanton term $\lambda$ at RK-DQCP.

\begin{figure}[t]
    \centering
    \includegraphics[width=1\linewidth]{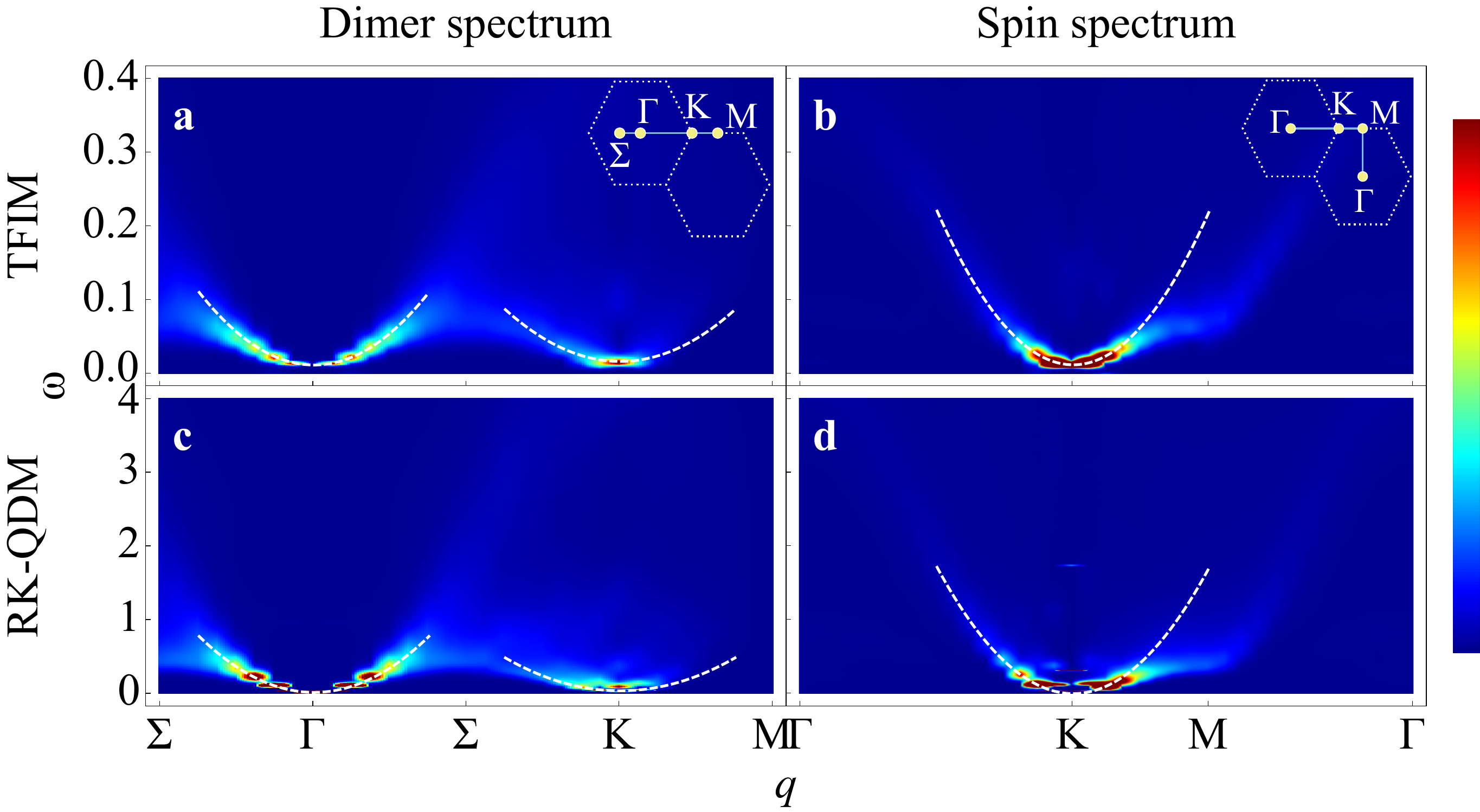}
    \caption{(a,c) Dimer-dimer and (b,d) spin-spin excitation spectra of the (a,b) $J_1$-$J_2$-$J_3$ TFIM at the multicritical point and (c,d) RK-QDM at RK-DQCP $v=t=1$. The spectra are measured within the $f=0$ topological sector and the system size is $L=24$.}
    \label{fig:4}
\end{figure}

\emph{Excitations.}---Here we turn to the dynamical properties at the multicritical point. We measure the dynamical dimer-dimer and spin-spin correlators defined as
\begin{equation}
    \begin{aligned}
        G_d (\mathbf{q},\tau)&=\frac{1}{L^2}\sum_\mathbf{RR'} e^{i(\mathbf{R}-\mathbf{R'})\cdot \mathbf{q}} \langle n^x _\mathbf {R} (0) n^x_\mathbf {R'} (\tau)\rangle,\\
        G_s (\mathbf{q},\tau)&=\frac{1}{L^2}\sum_{ij} e^{i(\mathbf{r}_i-\mathbf{r}_j)\cdot \mathbf{q}} \langle S_i^z (0) S_j^z (\tau)\rangle.
    \end{aligned}
\end{equation}
Here $n^x_\mathbf{R}\equiv 2S^z_{\mathbf{R}-\hat{x}/2} S^z_{\mathbf{R}+\hat{x}/2}+1$ indicates the density of the dimer centered at $\mathbf R$ along $x$-direction. As for comparison, we have also measured the dynamical excitation spectra for the RK wavefunction of the RK-QDM using the diffusion Monte Carlo technique~\cite{dmc_1,dmc_2,note_supp}. The excitation spectra are obtained from the imaginary time correlations via the stochastic analytical continuation technique~\cite{sac_1,sac_2,sac_3,note_supp}. 
At RK-DQCP, two gapless quadratic excitations have been identified in the dimer-dimer spectra~\cite{rk_spec,rk_excitation}: the resonon mode around $\Gamma$ point corresponding to the gauge photon excitations, and the so-called `pi0n' excitation around $\mathrm{K}$ point due to the proximity to the plaquette ordered phase. In our numerical spectra, we indeed observe these two quadratic excitations~(Fig.~\ref{fig:4}a,c). The quadratic dispersion is consistent with the dynamical exponent $z=2$ at RK-DQCP. Also, these two excitations differ by curvatures~\cite{note_supp}, which excludes the possibility of band folding and indicates the distinct physical nature of these two excitations.

Then we turn to the spin-spin correlators. As the $S^z$ operator is odd under $\mathscr G$, the spin-spin correlator measures excitations that live \emph{out of} the dimer Hilbert space. Thus one has to be extremely careful assigning these excitations. In the spin-spin spectrum, we only observe one branch of quadratic excitation stemming from the $\mathrm{K}$ point. This excitation shares the same origin with the `pi0n' excitation in the QDM, as the proximate clock phase orders at $\mathrm{K}$ point. However, it differs from QDM `pi0n' excitation as it lives out of the dimer Hilbert space. We dub this new excitation as `pi0n*'. Further, the curvature of this pi0n* is different from the previous resonon and pi0n, which confirms a different nature of this excitation.

As the clock and stripe phases live in different topological sectors, the clock-to-stripe transition at RK-DQCP should be irrelevant to the condensation of any point-like particles. Indeed, from the spectra, we see that the spin gap at the $\mathrm{M}$ point remains finite across RK-DQCP, which confirms that the stripe phase is not developed through n\"aive quasi-particle condensation. Instead, it can be understood as the proliferation of topological string objects that bridges different topological sectors: dimer or spin configurations can alternatively be represented as nonintersecting strings that form closed loops on the lattic~\cite{note_supp}. The string tension vanish at RK-DQCP, which results in a `string liquid' state where the motion and proliferation of strings are free from energy costs. Meanwhile, the topological winding numbers $(F_{1},F_{2})$ are entirely contributed by the strings on non-contractible loops which we dub as `topological strings'. From the string perspective one can also understand the exact degeneracy of RK-DQCP among different  topological sectors from the absence of string tension at this point.

\emph{Intermediate regime.}---In the phase diagram, we find an interesting fan-shaped regime with intermediate `tilt' $f$ at the clock-to-stripe boundary. The static spin structural factor is peaked at some intermediate wave vectors between the K and M points depending on the value of $f$ in the Brillouin zone. The most exciting feature of this intermediate regime is that this regime meets the microscopic ingredients for the `Cantor deconfinement' scenario~\cite{rk_2dbp}, where the incommensurate states are interleaved with commensurate ones, filling a fractal structure of incomplete `devil's staircase'. More mathematically, the incommensurate phases form a generalised Cantor set with finite measure. The incommensurate states exhibit gapless phason mode, which acts as gapless gauge photon mediating Coulomb interaction between spinons, hence prevents the system from confining.

Due to the limited system size in our numerical measurement, we are unable to observe the fractal structure of this intermediate regime. A detailed analysis of this incommensurate regime is beyond the scope of our current work. However, we notice that the parameter line $J_3=J_2/2$ in our $J_1$-$J_2$-$J_3$ TFIM can be exactly mapped to the $v_0=v_3$ line in the extended QDM~\cite{note_supp,qdm_honey}, where the latter cuts through the intermediate regime and its properties have been elaborately discussed~\cite{qdm_honey}.

\emph{Discussions.}---The experimental realisation of RK points has been a challenging task until recently. To our knowledge, our generalised TFIM the first experimental relevant system to simulate QDM physics that accesses both the RK-DQCP and the `Cantor deconfinement' regime. The TFIM is a realistic model in a variety of contexts, such as rare-earth frustrated magnets~\cite{tmgo_cava,tmgo_neu,tmgo_uud,tmgo_kt,tmgo_model,tmgo_nmr,tmgo_mpdf,tmgo_field}, dielectric materials~\cite{chai}, cold atom systems~\cite{coldatom} and even superconducting qubits~\cite{sqbit}. In addition, the RK-DQCP that appears in our system exhibit clear experimental signatures. The characteristic feature of the quadratic excitations can be directly observed by probing dynamical correlations. In condensed matter experiments, the dynamical spin-spin correlations can be measured by inelastic neutrons, while the dimer-dimer ones can be accessed in optics such as inelastic X-ray measurements. Moreover, it is worth noting that the gauge photons that appear in the dimer-dimer spectra have pronounced bandwidth $\sim\Delta$, much more visible than those in other realistic systems with emergent gauge structures such as quantum spin ice systems~\cite{CJH_PRL}.

Here we discuss the feasibility of our proposal realizing the exotic RK-DQCP and the `Cantor deconfinement' regime in experiments. This regime lies in the parameter regime $J_{3}/J_{2}\sim 0.4$ in our phase diagram. In condensed matter systems, magnetic dipole-dipole interaction is a natural sourse that mediates long-ranged spin interaction with $r^{-3}$ power-law decay. Pure dipole-dipole interaction yields  $J_{3}/J_{2}\sim 2/3$ which deviates from our desired regime. However, the ratio $J_{3}/J_{2}$ can be mulnipulated considering superexchange mechanism. With proper material design, it is still possible to hit this desired regime in condensed matter experiments. A more flexible platform of quantum control is the ultra-cold atoms, especially recent fast-developing programmable Rydberg arrays which are straightforwardly described with TFIM~\cite{Rydberg_review20}. Their geometry can be flexible assembled, and the effective transverse filed can be tuned by changing the interaction between photons and atoms~\cite{Rydberg_array}. Although the dipole-dipole interaction between Rydberg atoms is usually power-law decaying type, the ratio of $J_1$-$J_2$-$J_3$ can still be fine-tuned if we consider Rydberg dressed atoms following soft-core potential~\cite{Rydberg_dress}. In realistic systems the effective model can contain other perturbations beyond our $J_1$-$J_2$-$J_3$ TFIM, so one may wonder if our proposal is stable against these perturbations. In fact, our proposal is based on a field-theoretical argument that rely on symmetries and is not sensitive to the microscopic details: RK-DQCP can be fine-tuned by two relevant perturbations as long as the symmetry is not broken. Moreover, the longer distance interactions can also be suppressed via Floquet engineering~\cite{Rydberg_array}.

\emph{Acknowledgements.}---We wish to thank Rong Yu, Yue Yu, Yiming Wang, Jiucai Wang and Chun-Jiong Huang for valuable discussions. C. L. thank Rong Yu for hospitality visiting Renmin University of China where part of the work is done. This work is supported by the National Key Research and Development Program of China (Grants Nos. 2017YFA0304204, 2016YFA0300501 and 2016YFA0300504), the National Natural Science Foundation of China (Grants Nos. 11625416 and 11474064), and the Shanghai Municipal Government (Grants Nos. 19XD1400700 and 19JC1412702). X.-F. Z. acknowledges funding from the National Science Foundation of China under Grants No. 11804034, No. 11874094 and No.12047564, Fundamental Research Funds for the Central Universities Grant No. 2020CDJQY-Z003.

\clearpage
\onecolumngrid
\fontsize{10pt}{15pt}\selectfont

\setcounter{page}{1}
\setcounter{equation}{0}
\setcounter{figure}{0}

\renewcommand{\thefigure}{S\arabic{figure}}
\renewcommand{\thepage}{S\arabic{page}}
\renewcommand{\theequation}{S\arabic{equation}}
\renewcommand{\thetable}{S\Roman{table}}

\subsection*{Emergent Rokhsar-Kivelson point in realistic quantum Ising models}

\begin{center}
    \noindent\large{\textbf{Supplemental Materials}}
\end{center}

\vspace{1mm}

\begin{enumerate}[label=\Roman*.]
    \item Height field representation and the topological flux
    \item Exact correspondence between TFIM and the extended QDM
    \item Finite size scaling of the multicritical point
    \item Spin-spin correlation function
    \item Emergent $U(1)$ symmetry
    \item Histogram of the clock order parameter across RK-QCP
    \item Linear-quadratic crossover
    \item Curvatures of quadratic dispersion
    \item The intermediate regime
    \item Stochastic series expansion (SSE)
    \item Stochastic analytical continuation (SAC)
\end{enumerate}

\section{Height field representation and the topological flux}

In this section, we give an introduction to the height field representation and the topological flux of the quantum dimer model. 

In the height representation, each hexagonal plaquette is assigned an integer number $h(\mathbf r)$. Turning clockwise around a site of the even, the height $h(\mathbf r)$ changes by $+1$ when crossing an empty link, and by $-2$ when crossing an occupied link. The height field ressembles the vector potential $\mathbf A$ in the $U(1)$ gauge theory. Correspondingly, the magnetic field $\mathbf B=\nabla\times\mathbf A$ is perpendicular to the slope. 

\begin{figure}[t!]
    \centering
    \includegraphics[width=0.3\linewidth]{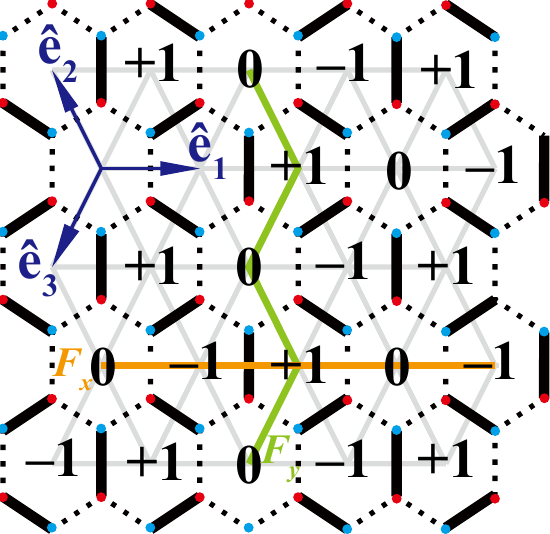}
    \caption{An illustration of the height field. The red and blue points denote the even and odd lattice site on the dual lattice. The numbers in the center of hexagonal plaquettes illustrate the height field of the corresponding dimer configuration. The dark blue arrows denote the three unit vectors defined on the triangular lattice and its dual honeycomb lattice of dimers. The light orange and green paths denote the loop along which the winding numbers $F_x$ and $F_y$ are defined. }
    \label{fig:s0}
\end{figure}

When defined on a torus, the total change of the height field after going a period around the boundary determines the topological sector. Such sectors can be labelled by a pair of flux quantum numbers $(F_x,F_y)$. Going around a chosen loop~(Fig.~\ref{fig:s0}), we add $2$ when crossing an occupied dimer and $-1$ when crossing an empty link. The flux quantum numbers divided by the system size $L_x$, $L_y$ are defined as the magnetic flux. 
\begin{equation}
    f_x=F_x/L_x,\quad f_y=F_y/L_y
\end{equation}
Such flux also corresponds to the tilt of the height field $h$. For the ground states, the `tilt' is locked along one of the three dimer directions $\hat{\mathbf{e}}_i$~($i=1,2,3$). Therefore, without loss of generality, we can set $f_y=0$ only work with $f\equiv f_x$. The clock (plaquette) phase corresponds to zero-tilt $f=0$, and the stripe (staggered) phase corresponds to the maximum tilt $f=2$.

\section{Exact correspondence between TFIM and the extended QDM}

For the $J_1$-$J_2$ TFIM, the transition from the clock to the stripe state is trivially first-order, suggesting $g_3>0$ at the transition. In the following we will establish the exact correspondence between our $J_1$-$J_2$-$J_3$ TFIM at $J_3=J_2/2$, and the extended QDM proposed in Ref. \cite{qdm_honey} at $v_0=v_3$. For the extended QDM, along the $v_0=v_3$ line the clock and stripe phases are sandwiched by a finite area of incommensurate states with intermediate flux. This indicates that  $g_3<0$ in the $J_3=J_2/2$ case. From the above we conclude that we can always find $J_3$ in the range $0<J_3<J_2/2$ where $g_3$ is fine-tuned to be zero, so that RK-QCP can be obtained at the clock-to-stripe transition.

The extended QDM proposed in Ref. \cite{qdm_honey} takes the form
\begin{equation}
H=v_0\hat{n}_0+v_3\hat{n}_3-t\sum_{\hexagon}(|\ha\rangle\langle\hb|+\mathrm{h.c.})
\end{equation}
where $\hat{n}_{\alpha}$ ($\alpha=0,1,2,3$) counts the total number of honeycomb plaquettes with $\alpha$ dimers, respectively. 

To to establish the mapping, let's consider a perturbation term above the nearest-neighbor TFIM on the triangular lattice
\begin{equation}
H'=J'\sum_i\left(\sum_{\langle ij \rangle _1}S_{j}^{z}\right)^{2}.
\end{equation}
On the one hand, this Hamiltonian can be decomposed into pairwise Ising interactions
\begin{equation}
    \begin{split}
        H' & =J'\sum_i\left(\sum_{\langle ij \rangle _1}(S_{i}^{z})^{2}+2\sum_{\langle ij \rangle _1 \langle ij' \rangle _1, j\neq j'}S_{i}^{z}S_{j}^{z}\right)\\
        & =2J'\sum_{\langle ij\rangle_{1}}S_{i}^{z}S_{j}^{z}+4J'\sum_{\langle ij\rangle_{2}}S_{i}^{z}S_{j}^{z}+2J'\sum_{\langle ij\rangle_{3}}S_{i}^{z}S_{j}^{z}+\mathrm{const.}
    \end{split}
    \label{eq:hpspin}
\end{equation}
The nearest neighbor terms is trivial that can be absorbed in $J_{1}$ and is ignored here. The main consequence of $H'$ is that it induces next-nearest- and third-nearest-neighbor interactions with $J_{2}=2J_{3}=4J'$. On the other hand, the term $\left(\sum_{\langle ij \rangle _1}S_{j}^{z}\right)^{2}=\left(\sum_{\langle ij \rangle _1}S_{i}^{z}S_{j}^{z}\right)^{2}$ is related to density of dimers in the plaquette centered at $\mathbf{r}_i$: for given $i$, $|\sum_{\langle ij \rangle _1}S_{j}^{z}|=3,2,1$ and $0$ corresponds to 0,1,2,3-dimer in that plaquette respectively. We have
\begin{equation}
    \begin{split}
        H'&=J'(0^2\hat{n}_3+1^2\hat{n}_2+2^2\hat{n}_1+3^2\hat{n}_0)\\
        &=2J'(\hat{n}_0+\hat{n}_3)+\mathrm{const.}
    \end{split}
    \label{eq:hpqdm}
\end{equation}
In the simplification we have used the dimer sum rule identity $\hat{n}_{0}+\hat{n}_{1}+\hat{n}_{2}+\hat{n}_{3}=N$ and $2\hat{n}_{0}+\hat{n}_{1}-\hat{n}_{3}=0$. Comparing \eqref{eq:hpspin} and \eqref{eq:hpqdm} we conclude that the our Ising spin model with $J_{2}=2J_{3}$ correspond to the extended QDM with $v_{0}=v_{3}=J_{2}/2$. 

A quantum Monte Carlo calculation verifies this correspondence. On the $J_2=2J_3$ line, an incommensurate phase region with the flux $0<f<2$ emerges at roughly $0.14<J_2<0.17$, or $0.69<J_2/\Delta<0.85$. On the other hand, the devil's stailcase on $V_0=V_3$ line locates at roughly $0.7<V_3/t<1$. The lower boundaries coincides very well, whereas the deviation of the upper boundary results from finite size effect. 

\begin{figure}[ht]
    \centering
    \includegraphics[height=0.3\linewidth]{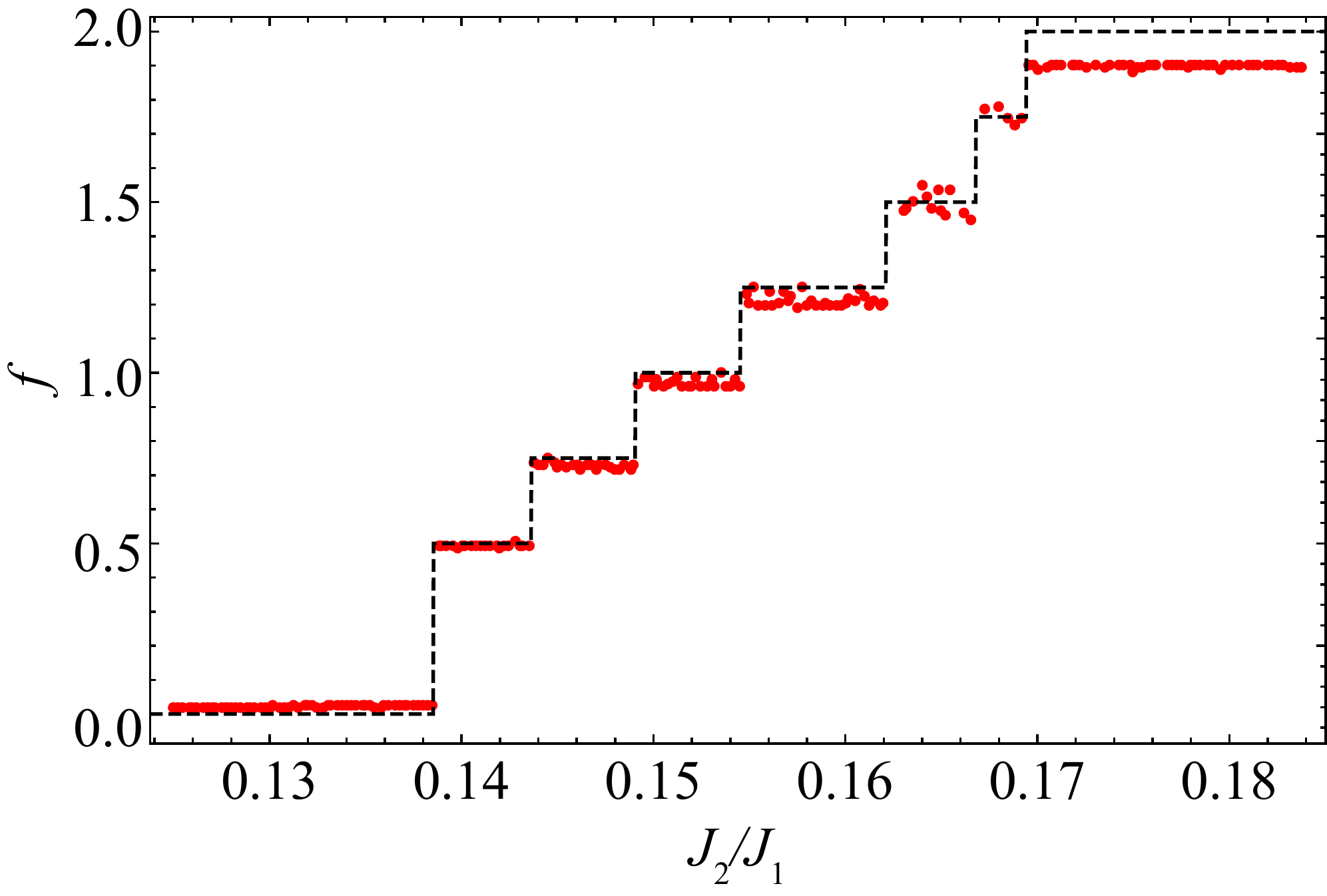}
    \includegraphics[height=0.3\linewidth]{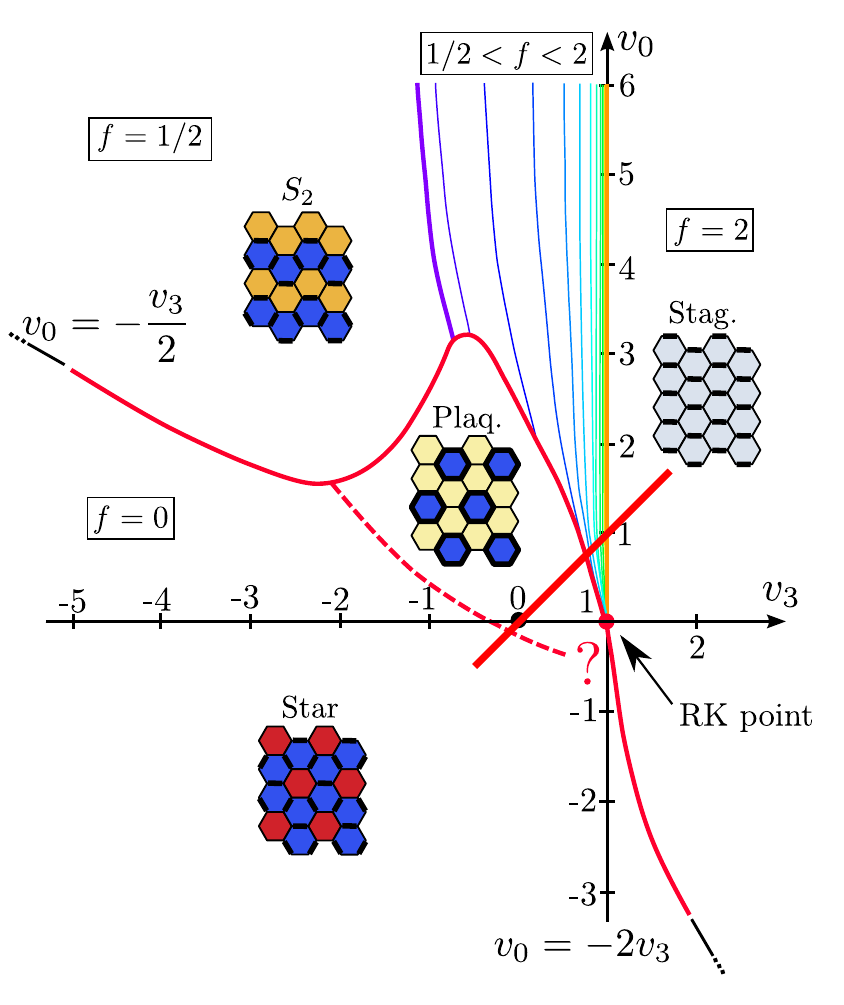}
    \caption{(a) The flux $f$ as a function of $J_2$ on the $J_2=2J_3$ line. $L=24$, $\beta=L$, $\Delta=0.2$ are taken. (b) The phase diagram of extendend dimer model taken from Ref. \cite{qdm_honey}. The $v_0=v_3$ line is highlighted. }
    \label{fig:s1}
\end{figure}

\section{Finite size scaling of the multicritical point}

\begin{figure}[ht]
    \centering
    \includegraphics[width=0.4\linewidth]{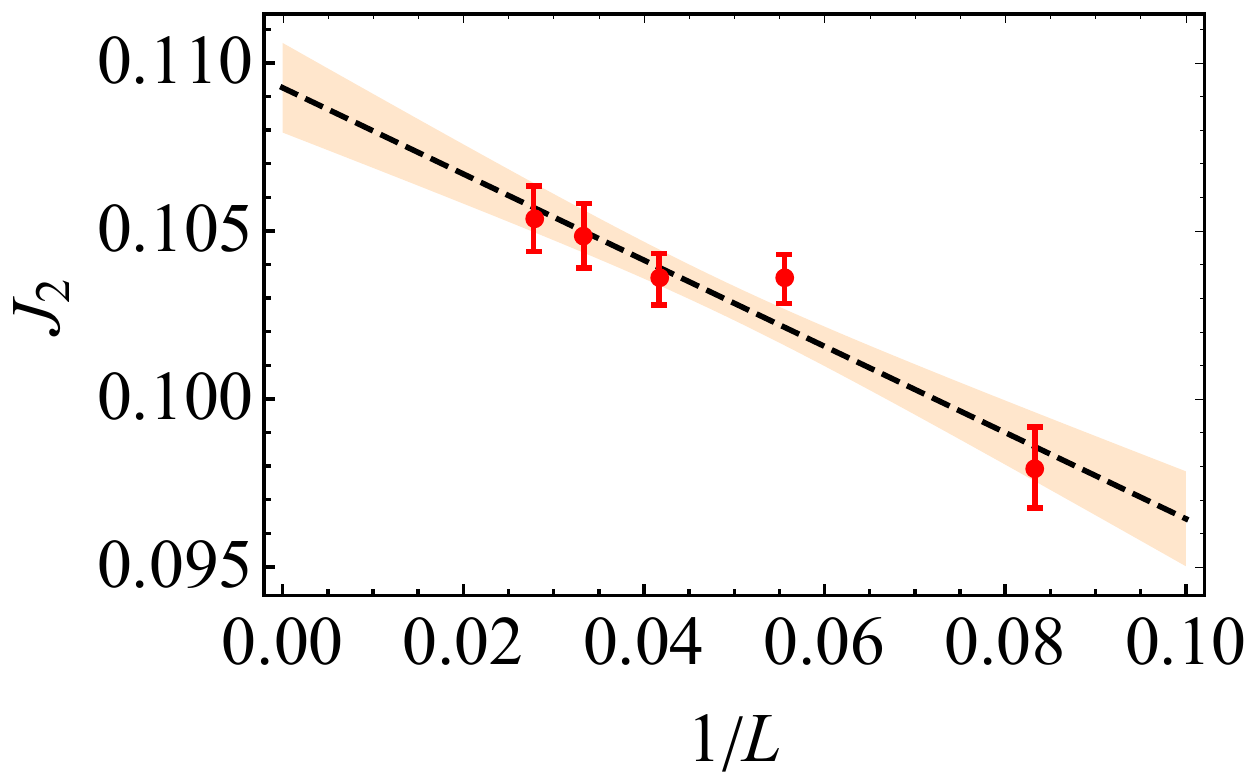}
    \includegraphics[width=0.4\linewidth]{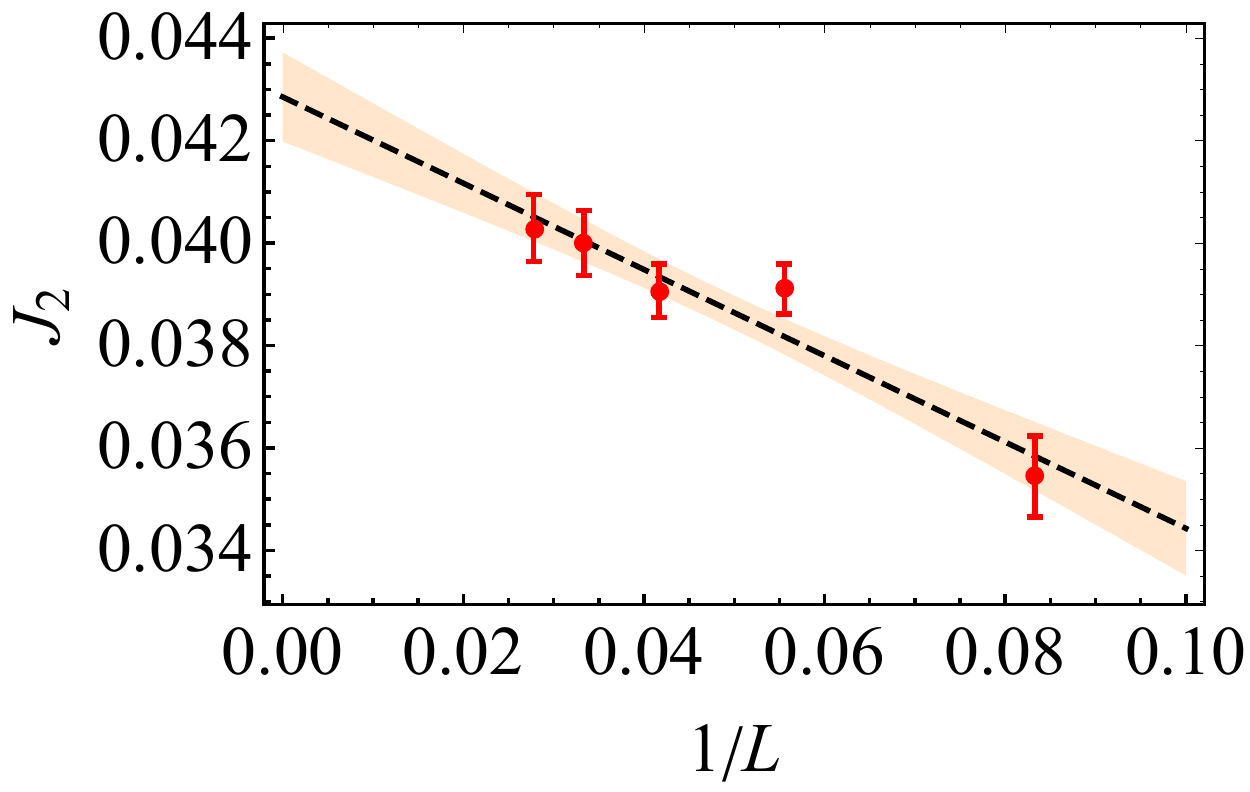}
    \caption{The finite size scaling which gives the position of the RK point in the thermodynamic limit}
    \label{fig:s5}
\end{figure}

The phase diagram in the main text is calculated at finite size $L=24$. To calculate the movement of the RK point when $L\rightarrow\infty$, we carried out a finite size scaling. The RK point is determined by determining the intersection of clock-stripe transition line and the stripe-incommensurate critical line, \textit{i.e.} $E(f=0)=E(f=2)=E(f=2-3/L_x)$. Through an extrapolation, we determine the position of the RK point in the thermaldynamic limit, which is $J_2=0.1093(11)$, $J_3=0.0429(7)$ (Fig. \ref{fig:s5}). 

\section{Spin-spin correlation function}

\begin{figure}[ht]
    \includegraphics[width=.6\linewidth]{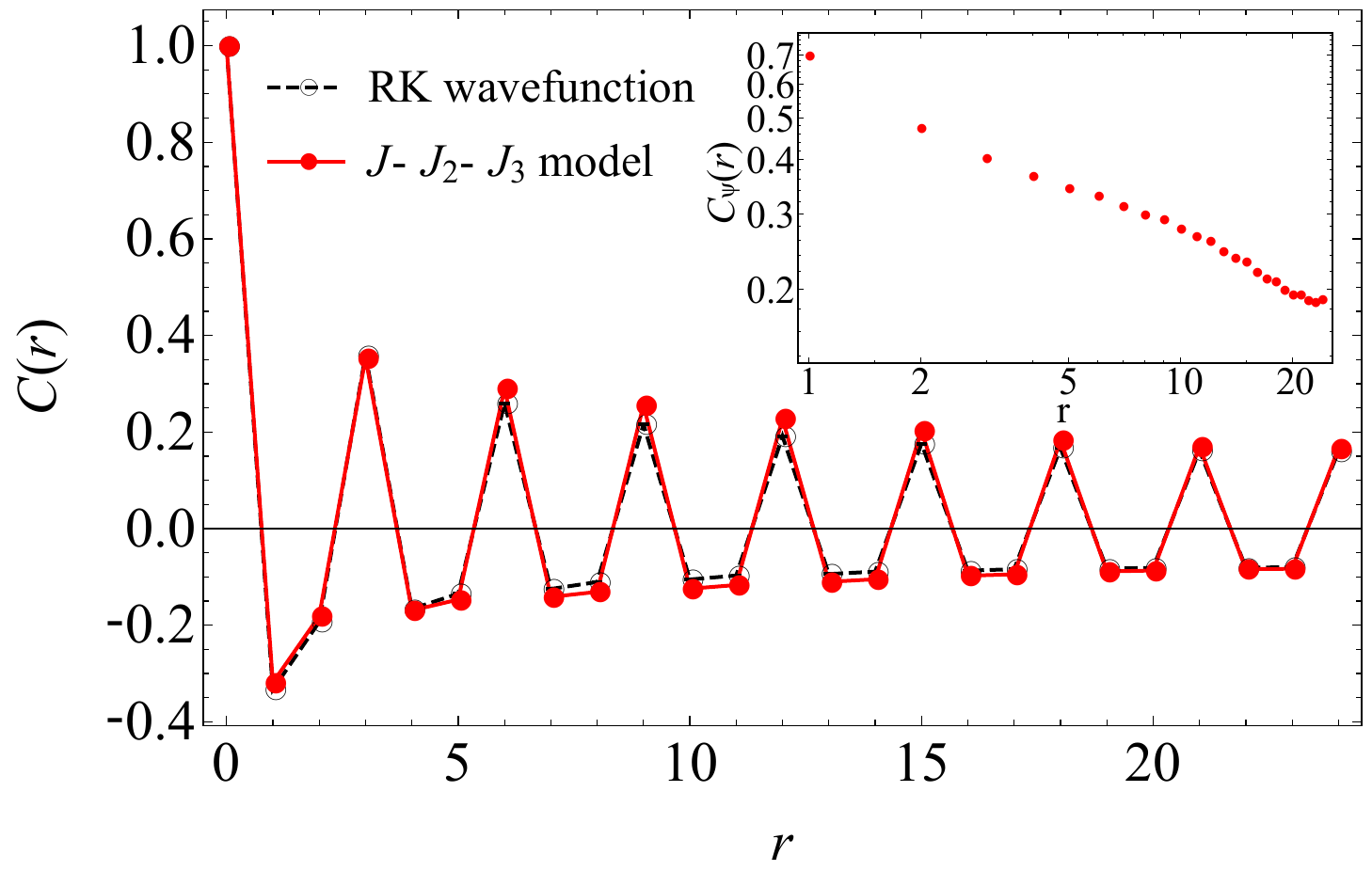}
    \caption{The spin-spin correlation function along $x$-axis, calculated for both $J$-$J_2$-$J_3$ model at the RK point and for tricritical wavefunction. Inset: the $U(1)$ spin correlation along $x$-axis in $J$-$J_2$-$J_3$ model at the RK point plotted in double-logarithm scale. Parameters $L=48$, $\beta=L^2$, $\Delta=0.2$ are taken for $J$-$J_2$-$J_3$ model and $L=48$ is taken for RK wavefunction}
    \label{fig:s2}
\end{figure}

We calculate the $S^z$-$S^z$ correlation at the multicritical point (Fig. \ref{fig:s2}). The correlations are measured along $x$-direction. Due to the fact that the RK wavefunction is the superposition of all the possible classical configurations, we can simultate the result of RK wavefunction by carrying out a classical Monte Carlo simulation of honeycomb-lattice dimer model at $\beta=0$. As cross-sector update is inaccesible in TFIM QMC, non-contractable loop updates are also banned in classical Monte Carlo simulation. These two correlation functions agrees considerably well, which confirms the RK property of the Ising tricritical point. 

However, as the correlation function oscillates as well as decays with the distance $r$, it's hard to determine the decaying behavior. So we need to turn to a correlation which decays monotonously, namely, the correlation of $U(1)$ spins. The $U(1)$ spin is defined on each triangle as a superposition of the $\mathbb{Z}_2$ spins on three sublattices:
\begin{equation}
    \psi=S_A^z+S_B^z\mathrm e^{\mathrm i2\pi/3}+S_C^z\mathrm e^{\mathrm i4\pi/3}
    \label{eq:clock}
\end{equation}
Such spin remains the same in single clock domain. Its correlation shows a algebraic decay behavior, which further confirms the RK nature of the Ising tricritical point. 

\section{Assymptotic behaviour of the correlators of RK wavefunction}

Here we evaluate the behaviour of the correlation functions $\langle n^x(0)n^x(\mathbf{r})\rangle$ and $\langle S^z(0)S^z(\mathbf{r})\rangle$ of RKQCP at large distance $|r|\rightarrow\infty$. The RK wave function is the equal weight superposition of classical dimer coverings $|\psi_\mathrm{RK}\rangle\sim\sum_{c}|c\rangle$. The observable with respect to this wavefunction $\langle\psi_\mathrm{RK}|\hat{\mathscr{O}}|\psi_\mathrm{RK}\rangle$ is equivalent to the statistical problem in classical dimer model at infinite temperature $T=\infty$, where the statistical weight of all dimer coverings are identical. 

For the dimer-dimer correlation function $G_d(\mathbf{R})=\langle n^x(0)n^x(\mathbf{R})\rangle$, such statistical problem has already been solved in Ref.~\cite{fisher1963statistical}, and the assymptotic behaviour is predicted to be $G_d(\mathbf{R})\sim|\mathbf{R}|^{-2}$. To measure the spin-spin correlation function $\langle S_i^zS_j^z\rangle$, we can construct a non-intersecting string $\mathscr{C}_{ij}=\{i,i_2,...,i_{N-1},j\}$ along nearest-neighbor triangular lattice bonds that terminates at $\mathbf{r}_i$ and $\mathbf{r}_j$. The spin-spin correlator $G_s(\mathbf{r}_i-\mathbf{r}_j)=\langle{S}_i^z{S}_j^z\rangle=2^{2-N}\langle({S}_i^z{S}_{i_2}^z)({S}_{i_2}^z{S}_{i_3}^z)\dots({S}_{i_{N-1}}^z{S}_j^z)\rangle$ can then be expressed in terms of density of dimers across the string, i.e., $\langle{S}_i^z{S}_j^z\rangle\sim\langle n^x_{ii_2}n^x_{i_2i_3}\dots n^x_{i_{N-1}j}\rangle$, where $n^x_{kl}$ denotes the dimer density on the link crossing the bond $\langle kl\rangle$. The assymptotic behaviour of such correlator is then predicted to be $G_s(\mathbf{r}_i-\mathbf{r}_j)\sim|\mathbf{r}_i-\mathbf{r}_j|^{-0.5}$.

\section{Emergent $U(1)$ symmetry}

\begin{figure}
    \centering
    \includegraphics[width=.4\linewidth]{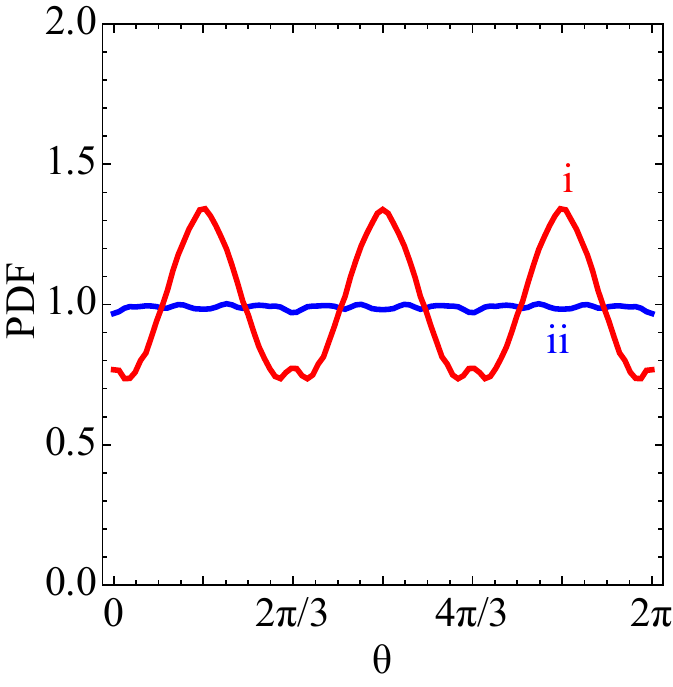}
    \caption{The angular probability density function (PDF) at point (i) in clock phase and (ii) at RK-DQCP. The system size of measurement is $L=24$}
    \label{fig:s8}
\end{figure}

To further confirm the emergent $U(1)$ symmetry, we measure the angular distribution of the VBS order parameter at the RK point and in the clock phase. We find that whereas in the clock phase, such distribution has three peaks with a $2\pi/3$ angle, at RK point, such distribution has no observable angular dependence, thus confirming the emergent $U(1)$ symmetry~(Fig.~\ref{fig:s8}). 

\section{Histogram of the clock order parameter across RK-QCP}

\begin{figure}[ht]
    \centering
    \includegraphics[width=.8\linewidth]{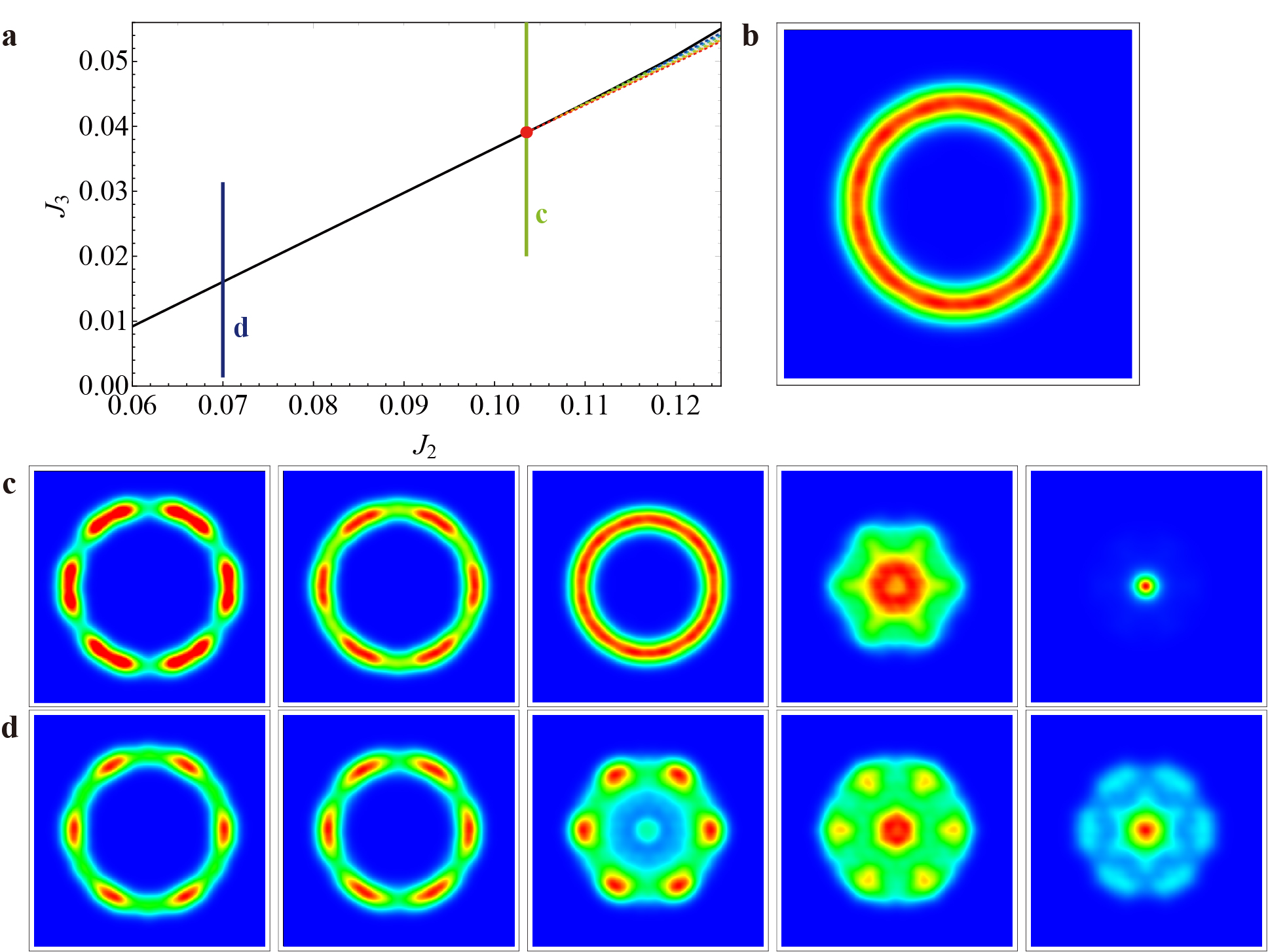}
    \caption{(a) An illustration of the three lines on the phase diagram we scan. (b) The histogram with $U(1)$ symmetry at RK point. $L=24$, $\beta=L^2/2=288$, $\rho_s=2/3$ is taken throughout the calculation of histograms. (c) The histogram of $U(1)$ order parameter crossing the RK point, $J_2=0.10356$, $J_3=0.080,0.060,0.039(\mathrm{RK}),0.030,0.020$  respectively. (d) The histogram of $U(1)$ order parameter along the transition line, $J_2=0.08,0.09,0.10,0.10356(\mathrm{RK}),0.11$ respectively. (e) The histogram of $U(1)$ order parameter in the first order region, $J_2=0.07$, $J_3=0.014,0.008,0.006,0.005,0.004$ respectively.}
    \label{fig:s3}
\end{figure}

Another important quantity to evaluate the phase transition is the $U(1)$ order parameter defined in \ref{eq:clock}. Since QMC update cannot cross topological sectors, we restrict ourselves to the $f=0$ sector, in which stripe order is substituted by a quasi-long range order. 

The clock phase, due to its 6-fold degeneracy, is characterised by six distinct peaks connected by a $C_6$ rotational symmetry, whereas a centralized peak is seen in the disordered phase. When we go from the clock phase to the disordered phase, the six peaks shrinks into one centralized peak (Fig. \ref{fig:s3}, line 2). 

The distinction between first order transition and RK-type transition is that at the first order transition point, the $\mathbb Z_6$ angular dependence pertains, whereas at the RK point, the angular dependence of the $\mathbb Z_6$ peaks gradually fade away, and a $U(1)$ symmetry emerges, characterised by a ring in the histogram (Fig. \ref{fig:s3}b) \cite{yan_sweeping}. This can be seen clearly numerically that when we go along the transition line from the first order end to the RK point, the $\mathbb Z_6$ feature gradually disappears and the six peaks connect into one ring, and after crossing the RK point, the $U(1)$ symmetry is broken (Fig. \ref{fig:s3}, line 3). 

By contrast, in the first order region, when we enter the disordered region, central peak and $\mathbb Z_6$ peaks are detected simultaneously (Fig. \ref{fig:s3}, line 4), and the weight transfers from the $C_6$ peaks to the central peak. When we cross the transition line, central peak isn't detected immediately, as the clock phase is still a metastable state. Double peak only occurs when such metastable state collapses, in a region deep inside the disordered phase. 

\section{Linear-quadratic crossover}

\begin{figure}[ht]
    \centering
    \includegraphics[width=0.5\linewidth]{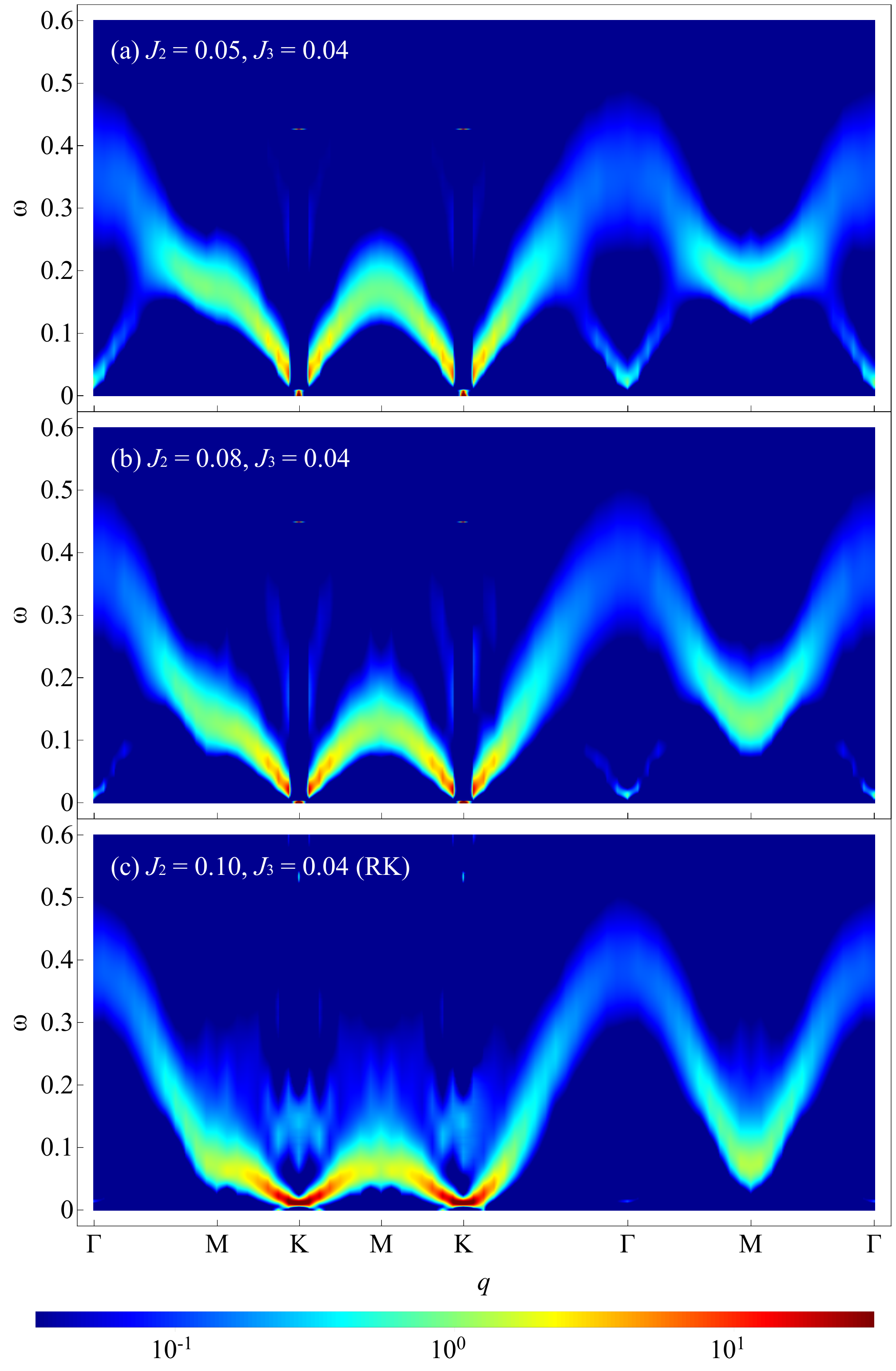}
    \caption{The linear-quadratic crossover in spin spectrum, measured at (a) $J_2=0.05$, (b) $J_2=0.08$, (c) $J_2=0.1035$ (RK point) and $J_3=0.0391$. $L=24$, $\beta=L^2/2=288$, $\rho_s=2/3$ is taken. }
    \label{fig:s4}
\end{figure}

When far away from the RK point, two linear modes are found at $\Gamma$ and $K$. When approaching the RK point, the nature of the effective gauge field changes from $(2+1)$d to $(2+2)$d. In the linear-quadratic crossover process (Fig. \ref{fig:s4}), the linear mode at $K$ softens into a quadratic mode, whereas the mode at $\Gamma$ gradually vanishes. 

\section{Curvatures of quadratic dispersions}

To explore the relation between the mode in spin correlation spectrum and the two modes in dimer correlation spectrum, we extracted the peak of intensity at different momentum and fit them to the form 
\begin{equation}
    \omega=\frac{1}{2}C_2(\mathbf q-\mathbf q_0)^2
    \label{eq:c2}
\end{equation}
in vicinity of the gapless point $\mathbf q_0$. The results are shown in Table~\ref{tbl:1}. We find the curvature of the pi0n* mode in spin spectrum roughly twice the curvature of pi0n in dimer spectrum.

\begin{table}[ht]
    \caption{The curvature defined in Eq. \ref{eq:c2} for different excitations measured in $J_1$-$J_2$-$J_3$ TFIM and RK-QDM. }
    \begin{tabular}{cc|cc}
        \hline\hline
        &&\multicolumn{2}{c}{Curvature $C_2$}\\
        \multicolumn{2}{c|}{Mode}&$J_1$-$J_2$-$J_3$ TFIM&RK-QDM\\
        \hline
        (\emph{Dimer spectrum})&Resonon&$0.080(3)$&$0.61(3)$\\
        &Pi0n&$0.057(4)$&$0.36(4)$\\
        (\emph{Spin spectrum})&Pi0n*&$0.095(2)$&$0.78(8)$\\
        \hline\hline
    \end{tabular}
    \label{tbl:1}
\end{table}

\section{The intermediate regime}

\begin{figure}
    \centering
    \includegraphics[width=0.6\linewidth]{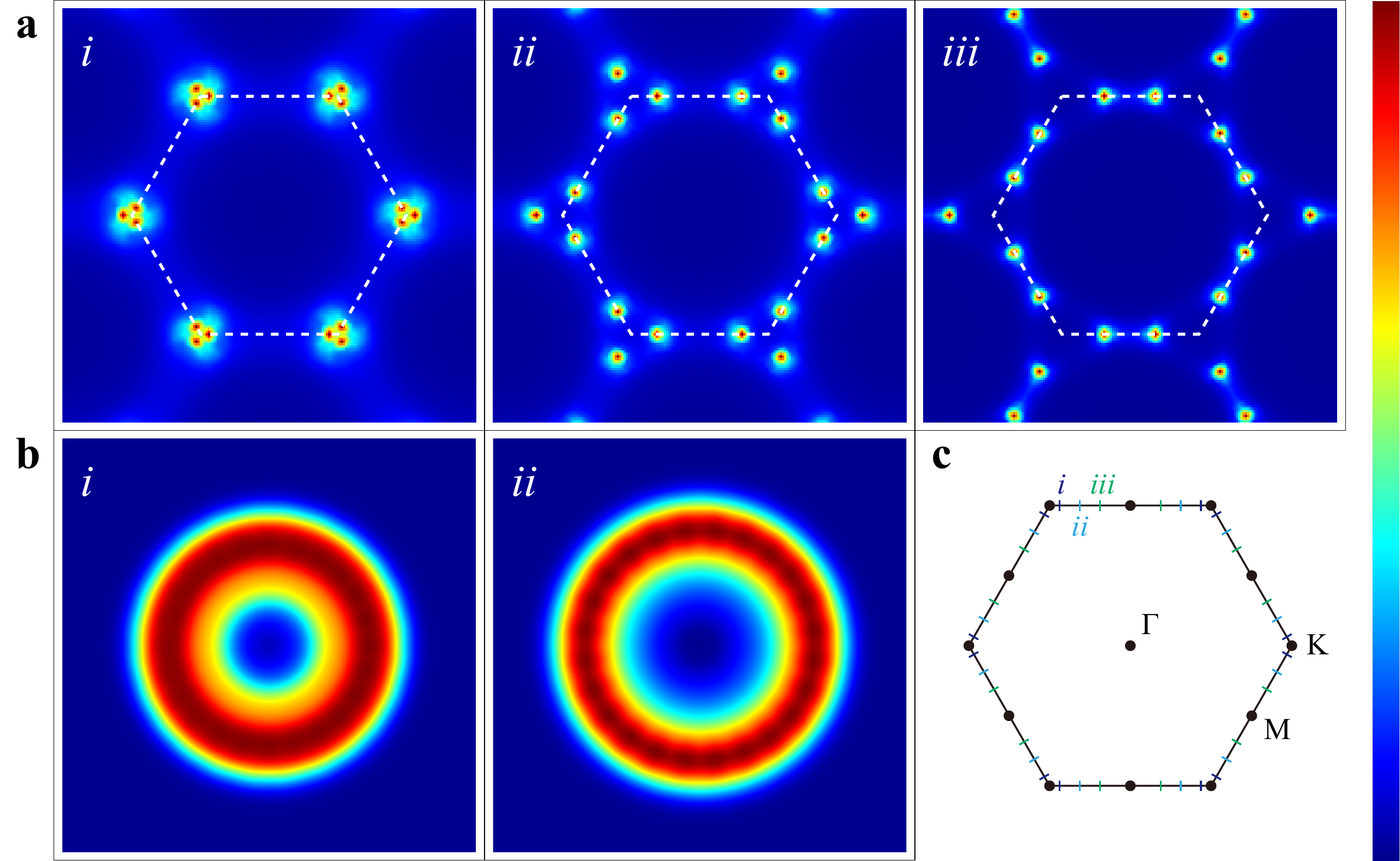}
    \caption{(a) The spin structure factor and (b) the complex order parameter distribution in the incommensurate phase, measured at $J_2=0.20$ and (i) $J_3=0.113$ where $f=0.25$, (ii) $J_3=0.107$ where $f=0.75$ and (iii) $J_3=0.102$ where $f=1.25$. (c) An illustration of the structure factor peaks in the Brillouin zone.}
    \label{fig:s7}
\end{figure}

We measure the spin structure factor
\begin{equation}
    S(\mathbf{q})=\sum_{ij}\langle S_i^zS_j^z\rangle\mathrm{e}^{\mathrm{i}\mathbf{q}\cdot(\mathbf{r}_i-\mathbf{r}_j)}
\end{equation}
in the incommensurate phase. We fix $J_2=0.20$ and vary $J_3$ so that the system can be ordered at different momentum points. At sufficiently large $J_3$, the clock phase is stabilized. When gradually decrease $J_3$, we find the peak at $\mathrm{K}$ point splits into three peaks and gradually move towards $\mathrm{M}$ point~(Fig.~\ref{fig:s7}). The location of the peaks is connected with the flux of the corresponding ordered phase by 
\begin{equation}
    \mathbf{q}=\left(\pm\frac{2(1-2f)\pi}{3},\pm\frac{2\pi}{\sqrt{3}}\right),\left(\pm\frac{(4+f)\pi}{6},\pm\frac{(2-f)\pi}{\sqrt{3}}\right),\left(\pm\frac{(8-f)\pi}{6},\pm\frac{f\pi}{\sqrt{3}}\right).
\end{equation}
This is a clear hallmark of an intermediate ordered phase with intermediate tilt $0<f<2$. 

Such intermediate phase can be characterised by a $U(1)$ order parameter
\begin{equation}
    \psi_\mathbf{q}=\sum_{i}S_i\mathrm{e}^{\mathrm{i}\mathbf{q}\cdot\mathbf{r}_i},
\end{equation}
where $\mathbf{q}$ is chosen to be the momentum point at which the system is ordered. We find such order parameter distributes evenly on a ring~(Fig.~\ref{fig:s7}b). This is an evidence for the incommensurate nature of the order. The $U(1)$ symmetry emerges because when one translate the system by one unit along the $\mathbf{x}$ direction, the order parameter gains an extra incommensurate phase factor,
\begin{equation}
    \mathbf{r}\to\mathbf{r}+\hat{\mathbf{x}},\quad \psi\to\psi\mathrm{e}^{\mathrm{i}q_x}
\end{equation}

\section{Stochastic series expansion (SSE)}

For the numerical works in this paper, we use a quantum Monte Carlo (QMC) method with stochastic series expansion (SSE) algorithm \cite{qmc_1,qmc_2,qmc_3} to calculate the ground state properties and imaginary time Green function. This method will be briefly introduced below. 

In quantum statistics, the measurement of observables is closely related to the calculation of partition function $Z$
\begin{equation}
    \langle\mathscr{O}\rangle=\mathrm{tr}\:\left(\mathscr{O}\exp(-\beta H)\right)/Z,\quad Z=\mathrm{tr}\exp(-\beta H)
\end{equation}
where $\beta=1/T$ is the inverse temperature, $H$ is the Hamiltonian of the system and $\mathscr{O}$ is an arbitrary observable. Typically, in order to evaluate the ground state property, one takes a sufficiently large $\beta$ such that $\beta\sim L^z$, where $L$ is the system scale and $z$ is the dynamical exponent. In SSE, such evaluation of $Z$ is done by a Taylor expansion of the exponential and the trace is taken by summing over a complete set of suitably-chosen basis. 
\begin{equation}
    Z=\sum_\alpha\sum_{n=0}^\infty\frac{\beta^n}{n!}\langle\alpha|(-H)^n|\alpha\rangle
\end{equation}
We then write the Hamiltonian as the sum of a set of operators whose matrix elements are easy to calculate. 
\begin{equation}
    H=-\sum_iH_i
\end{equation}
In practice we truncate the Taylor expansion at a sufficiently large cutoff $M$ and it is convenient to fix the sequence length by introducing in identity operator $H_0=1$ to fill in all the empty positions despite it is not part of the Hamiltonian. 
\begin{equation}
    (-H)^n=\sum_{\{i_p\}}\prod_{p=1}^nH_{i_p}=\sum_{\{i_p\}}\frac{(M-n)!n!}{M!}\prod_{p=1}^nH_{i_p}
\end{equation}
and 
\begin{equation}
    Z=\sum_\alpha\sum_{\{i_p\}}\beta^n\frac{(M-n)!}{M!}\langle\alpha|\prod_{p=1}^nH_{i_p}|\alpha\rangle
\end{equation}

To the carry out the summation, a Monte Carlo procedure can be used to sample the operator sequence $\{i_p\}$ and the trial state $\alpha$ with according to their relative weight
\begin{equation}
    W(\alpha,\{i_p\})=\beta^n\frac{(M-n)!}{M!}\langle\alpha|\prod_{p=1}^nH_{i_p}|\alpha\rangle
\end{equation}
For sampling we adoopt a Metropolis algorithm where the configuration of one step is generated based on updating the configuration of the former step and the update is accepted at a probability
\begin{equation}
    P(\alpha,\{i_p\}\rightarrow\alpha',\{i'_p\})=\min\left(1,\frac{W(\alpha',\{i'_p\})}{W(\alpha,\{i_p\})}\right)
\end{equation}
Diagonal update, where diagonal operators are inserted into and removed from the operator sequence, and cluster update, where diagonal and off-diagonal operates convert into each other, are adopted in update strategy. 

\begin{figure}[b]
    \centering
    \includegraphics[width=0.6\linewidth]{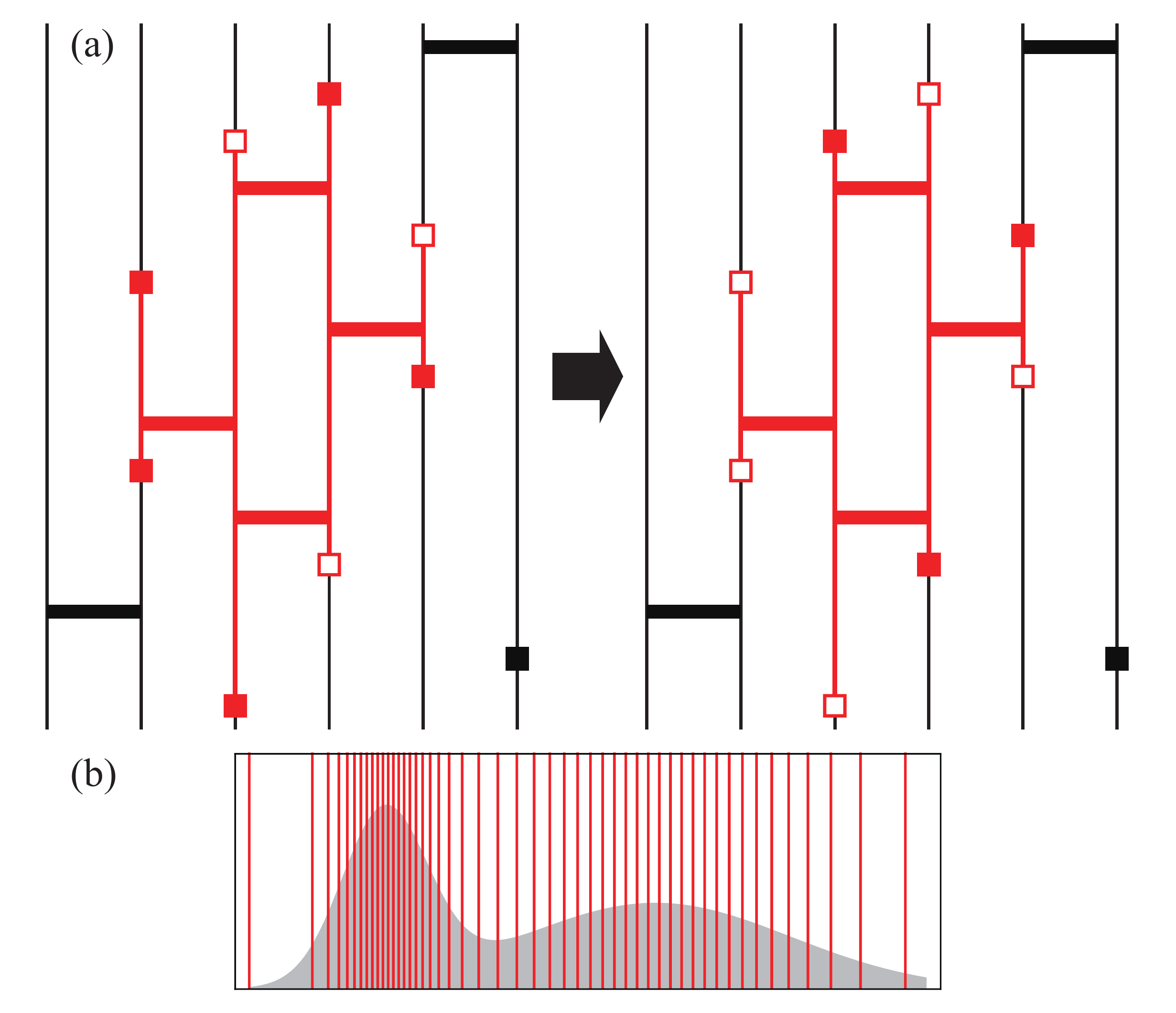}    
    \caption{(a) An illustration of the SSE cluster update process where the cluster marked in red is identified and filpped as a whole. A vertical line shows a spin expanded. The solid and empty squares show the diagonal and off-diagonal site operators. The solid bars show the diagonal bond operators. (b) An illustration of the parametrization of a continuous spectral function into discrete $\delta$-functions. }
    \label{fig:s6}
\end{figure}

In transverse field Ising model $H=J\sum_bS_{i_b}^zS_{j_b}^z-h\sum_i\sigma_i^x$, we write the Hamiltonian as the sum of following operators
\begin{equation}
    \begin{aligned}
        H_0&=1\\
        H_i&=h(S_i^++S_i^-)/2\\
        H_{i+n}&=h/2\\
        H_{b+2n}&=J(1/4-S_{i_b}^zS_{j_b}^z)
    \end{aligned}
\end{equation}
where a constant is added into the Hamiltonian for convenience. For the non-local update, a branching cluster update strategy is constructed \cite{qmc_2}, where a cluster is formed in $(D+1)$-dimensional by grouping spins and operators togather. Each cluster terminates on site operators and includes bond operators (Fig. \ref{fig:s6}a). All the spins in each cluster is flipped together at a probability $1/2$ agter all clusters are identified. 

\section{Stochastic analytical continuation (SAC)}

For the spectra in this paper we adopted a stochastic analytical continuation (SAC) \cite{sac_1,sac_2,sac_3} method to obtain the spectral function $S(\omega)$ from the imaginary time correlation $G(\tau)$ measured from QMC, which is generally believed a numerically unstable problem. This method will be briefly introduced below. 

The spectral function $S(\omega)$ is connected to the imaginary time Green's function $G(\tau)$ through an integral equation 
\begin{equation}
    G(\tau)=\int_{-\infty}^{\infty}\mathrm{d}\omega S(\omega)K(\tau,\omega)
\end{equation}
where $K(\tau,\omega)$ is the kernal function depending on the temperature and the statistics of the particles. We restrict ourselves to the case of spin systems and with only positive frequencies in the spectral, where $K(\tau,\omega)=(e^{-\tau\omega}+e^{-(\beta-\tau)\omega})/\pi$. To ensure the normalization of spectral function, we further modity the transformation and come to the following equation :
\begin{equation}
    G(\tau)=\int_0^\infty\frac{\mathrm{d}\omega}{\pi}\frac{e^{-\tau\omega}+e^{-(\beta-\tau)\omega}}{1+e^{-\beta\omega}}B(\omega)
    \label{eq:kernal}
\end{equation}
where $B(\omega)=S(\omega)(1+e^{-\beta\omega})$ is the renormalized spectral function. 

In practice, $G(\tau)$ for a set of imaginary time $\tau_i(i=1,\cdots N_\tau)$ is measured in QMC simultation together with the statistical errors. The renormalized spectral function is parametrized into large number of equal-amplitude $\delta$-functions whose positions are sampled (Fig. \ref{fig:s6}b)
\begin{equation}
    B(\omega)=\sum_{i=0}^{N_\omega}a_i\delta(\omega-\omega_i)
\end{equation}
Then the fitted Green's functions $\tilde{G}_i$ from Eq. \ref{eq:kernal} and the measured Greens functions $\bar{G}_i$ are compared by the fitting goodness
\begin{equation}
    \chi^2=\sum_{i,j=1}^{N_\tau}(\tilde{G}_i-\bar{G}_i)(C^{-1})_{ij}(\tilde{G}_j-\bar{G}_j)
\end{equation}
where the covariance matrix is defined as 
\begin{equation}
    C_{ij}=\frac{1}{N_B(N_B-1)}\sum_{b=1}^{N_B}(G_i^b-\bar{G}_i)(G_j^b-\bar{G}_j),
\end{equation}
with $N_B$ the number of bins, the measured Green's functions of each $G_i^b$. 

A Metropolis process is utilized to update the series in sampling. The weight for a given spectrum is taken to follow a Boltzmann distribution
\begin{equation}
    W(\{a_i,\omega_i\})\sim\exp\left(-\frac{\chi^2}{2\Theta}\right)
\end{equation}
with $\Theta$ a virtue temperature to balance the goodness of fitting $\chi^2$ and the smoothness of the spectral function. All the spectral functions of sampled series $\{a_i,\omega_i\}$ is then averaged to obtain the spectrum as the final result. 


\begin{thebibliography}{73}
\bibitem [{not({\natexlab{c}})}]{note_3}%
  \BibitemOpen
  \href@noop {} {\bibfield  {journal} {\bibinfo  {journal} {In usual cases, the
  energy difference between topological sectors $\propto\Delta L^{2}$, where
  $\Delta$ is the typical energy scale of the effective QDM}}
  {\natexlab{c}}}\BibitemShut {NoStop}%
\end{thebibliography}
\end{document}